\documentclass[apj,twocolumn]{emulateapj}

\shorttitle{Assembly of Massive Galaxies in a High-z Protocluster}
\shortauthors{Uchimoto et al.}

\begin{document}

\title{Assembly of Massive Galaxies in a High-z Protocluster}

\author{YUKA K. UCHIMOTO \altaffilmark{1,2,3},
   TORU YAMADA \altaffilmark{1},
   MASARU KAJISAWA \altaffilmark{4},
   MARIKO KUBO \altaffilmark{1},
   TAKASHI ICHIKAWA \altaffilmark{1},
   YUICHI MATSUDA \altaffilmark{5},
   MASAYUKI AKIYAMA \altaffilmark{1},
   TOMOKI HAYASHINO \altaffilmark{6},
   MASAHIRO KONISHI \altaffilmark{2},
   TETSUO NISHIMURA \altaffilmark{3},
   KOJI OMATA \altaffilmark{3},
   RYUJI SUZUKI \altaffilmark{7,8},
   ICHI TANAKA \altaffilmark{3},
   CHIHIRO TOKOKU \altaffilmark{9},
   TOMOHIRO YOSHIKAWA \altaffilmark{10}
}
\email{uchimoto@astr.tohoku.ac.jp}

\altaffiltext{1}
{Astronomical Institute, Tohoku University, 6-3 Aoba, Aramaki,
Aoba-ku, Sendai, Miyagi, 980-8578, Japan}

\altaffiltext{2}
{Institute of Astronomy, University of Tokyo, 2-21-1 Osawa, Mitaka,
Tokyo, 181-0015, Japan}

\altaffiltext{3}
{Subaru Telescope, National Astronomical Observatory of
Japan, 650 North A'ohoku Place, 
Hilo, HI 96720, USA}

\altaffiltext{4}
{Research Center for Space and Cosmic Evolution, Ehime University, Bunkyo-cho 2-5, Matsuyama 790-8577, Japan}

\altaffiltext{5}
{Department of Physics, University of Durham, South Road, Durham DH1 3LE, UK}

\altaffiltext{6}
{Research Center for Neutrino Science, Graduate School of Science, 
Tohoku University, Aramaki, Aoba-ku, Sendai, Miyagi, 980-8578, Japan}

\altaffiltext{7}
{Thirty Meter Telescope Observatory Corp., 1200 E. California Blvd. Mail Code 102-8 Pasadena, CA 91125, USA}

\altaffiltext{8}
{National Astronomical Observatory of Japan, 2-21-1 Osawa, Mitaka, Tokyo 181-8588, Japan}

\altaffiltext{9}
{University of California, Riverside, 900 University Ave. Riverside, CA 92521, USA}

\altaffiltext{10}
{Koyama Astronomical Observatory, Kyoto Sangyo University, Motoyama, Kamigamo, Kita-ku, Kyoto 603-8555, Japan}

\begin{abstract}
 We present the results of wide-field deep $JHK$ imaging of the SSA22 field using MOIRCS instrument equipped with Subaru telescope. The observed field is 112 arcmin$^2$ in area, which covers the $z=3.1$ protocluster characterized by the overdensities of Ly$\alpha$ emitters (LAEs) and Ly$\alpha$ Blobs (LABs). The 5 $\sigma$ limiting magnitude is $K_{AB} = 24.3$. We extract the potential protocluster members from the $K$-selected sample by using the multi-band photometric-redshift selection as well as the simple color cut for distant red galaxies (DRGs; $J-K_{AB}>1.4$). The surface number density of DRGs in our observed fields shows clear excess compared with those in the blank fields, and the location of the densest area whose projected overdensity is twice the average coincides with the large-scale density peak of LAEs. We also found that $K$-band counterparts with $z_{\rm phot} \simeq 3.1$ are detected for 75 \% (15/20) of the LABs within their Ly$\alpha$ halo, and the 40 \% (8/20) of LABs have multiple components, which gives a direct evidence of the hierarchical multiple merging in galaxy formation. The stellar mass of LABs correlates with their luminosity, isophotal area, and the Ly$\alpha$ velocity widths, implying that the physical scale and the dynamical motion of Ly$\alpha$ emission are closely related to their previous star-formation activities. Highly dust-obscured galaxies such as hyper extremely red objects (HEROs; $J-K_{AB}>2.1$) and plausible $K$-band counterparts of submillimeter sources are also populated in the high density region.

\end{abstract}


\section{INTRODUCTION}

 Since the late 1990's, overdense regions of high redshift galaxies at $z\gtrsim2$ have been discovered and extensively investigated by many authors. The most notable examples are high density regions of Ly$\alpha$ emitters (LAEs), Lyman break galaxies (LBGs) and their resemblances over the redshift range of 2 - 5 (e.g. Steidel et al. 1998, 2000; Shimasaku et al. 2003; Ouchi et al. 2005). These studies revealed the properties of the on-going star formation activities in the high density regions of the universe at high redshift. However, the galaxies selected by the rest-frame ultra-violet (UV) emission may not be the only representative population in such dense regions. For example, sum of the stellar mass of a hundred of LAEs is only $10^{10-11}$ M$_{\odot}$, as the typical LAEs are known to be less massive objects with $\sim 10^{8-9}$ M$_{\odot}$ (Gawiser et al. 2007; Ono et al. 2010), which corresponds to only one typical massive galaxy.

 On the other hand, the overdensities around high-z radio galaxies and QSOs, which are expected to evolve into giant elliptical galaxies, are also observed so far (Kurk et al. 2004; Venemans et al. 2007; Overzier et al. 2009; Hatch et al. 2011). From recent near- to mid- infrared and submillimeter observations, the high-density regions of relatively mature galaxies or highly-obscured dusty starbursts are indeed detected around the radio galaxies, or X-ray selected cluster candidates (Kodama et al. 2007; Daddi et al. 2009; Gobat et al. 2011; Tanaka et al. 2011). The samples are still very limited, however, and the general properties of the protoclusters at high redshift are far from understood.

 The protocluster at $z=3.09$ in and around the SSA22 field was first discovered by Steidel et al. (1998, 2000) as a significant redshift peak of LBGs with the overdensity of $\delta_{\rm gal} \sim 5.0$ over the 20 Mpc-scale comoving volume. Later narrow-band observations have found the large-scale filamentary structure of LAEs extending over 60 Mpc (the SSA22-sb1 field; Hayashino et al. 2004). Yamada et al. (2012) revealed that the overdensity of LAEs is in fact the most prominent structure in the larger survey area of 1.38 deg$^2$, corresponding to the comoving volume of 10$^6$ Mpc$^3$.

 In this field, along the structure of LAEs, there also locate the extremely large, giant Ly$\alpha$ clouds, Ly$\alpha$-blob1 (hereafter LAB1) and blob2 (LAB2) (Steidel et al. 2000), and the other 33 LABs (Matsuda et al. 2004, 2007). The objects are considered to be closely related to the early phase of massive galaxy formation in the overdense regions at high redshift (e.g., Yamada 2009). LABs typically have the physical extent of $30-150$ kpc and the Ly$\alpha$ luminosity of $\gtrsim 10^{43}$ erg s$^{-1}$. They are mostly found in the high-density regions of star-forming galaxies at $z=2-5$ (Matsuda et al. 2004, 2011; Saito et al. 2008; Yang et al. 2010), while the most distant LAB is discovered by Ouchi et al. (2009) at $z=6.6$. The origin of the strong and extended Ly$\alpha$ emission is still unclear; it may be powered by cold gas accretion (Haiman et al. 200;  Goerdt et al. 2010a), or by galactic superwinds (Taniguchi \& Shioya 2000; Mori \& Umemura 2006), or by photoionization induced by intense starbursts or active galactic nuclei (AGN) (Chapman et al. 2001; Geach et al. 2009). Recent studies based on X-ray and mid-infrared observations suggests that a significant amount of LABs has an indication of AGN which can power the extended Ly$\alpha$ emission (Lehmer  et al. 2009; Geach et al. 2009; Webb et al. 2009). 

 \citet{key-uchimoto08} carried out a deep near-infrared observation in the southern part of the SSA22a field \citep{key-steidel00}. In the field (hereafter referred as SSA22-M1), they found density excesses of distant red galaxies (DRGs; Franx et al. 2003) and the photometric redshift (hereafter photo-z) selected objects in the vicinity of LAB1 and LAB2. Moreover, they showed that the stellar mass of K-selected objects associated with 8 LABs are in the range of $M_* = 10^9-10^{11}$ M$_{\odot}$, and they are correlated with their luminosity of Ly$\alpha$ emission. Since the observed field of view is not large enough to cover the entire high-density region of LAEs, however, the stellar mass distribution over the entire high-density structure was not clear very much.

 In this paper, we present the results of the new deep and wide near-infrared (NIR) imaging of the protocluster at $z=3.09$ in the SSA22-sb1 field by using {\it Multi-Object Infra-Red Camera and Spectrograph} (MOIRCS; Suzuki et al. 2008) mounted on 8.2 m Subaru Telescope. We here construct the large NIR-selected sample of protocluster galaxies down to $K=24.3$ along the filamentary, high-density structure of LAEs at $z=3.09$. The limiting magnitude corresponds to the stellar mass $\gtrsim 3 \times 10^{9}$ M$_{\odot}$ assuming the typical SED of LBGs at $z\sim 3$. We will focus on the stellar mass assembly in the protocluster at $z=3.1$ in this paper. Our scientific goal is to reveal how much of the stellar mass has already formed and assembled in the protocluster. In other words, we would like to see the relation between the structure traced by the star-formation activity and that traced by the stellar mass.

\begin{figure}[tbp]
\epsscale{1.0}
\includegraphics[width=8cm]{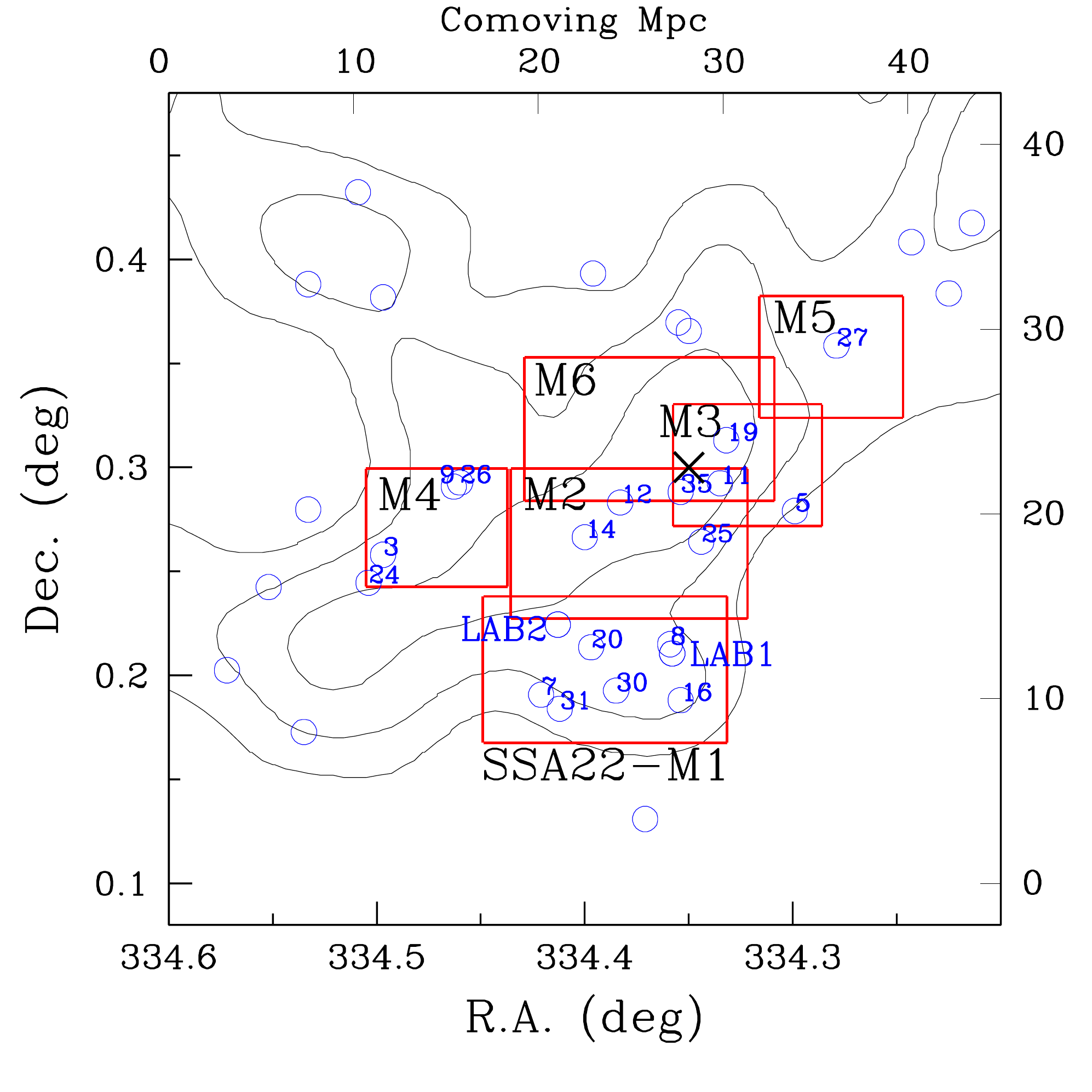}
\caption{
The observed area in the SSA22-Sb1 field. The rectangles (red) indicate each field of view, in which the names of the observed fields are also shown. Contours (black) are 1$\sigma$, 1.5$\sigma$, 2$\sigma$ density levels of LAEs. LABs are indicated with circles (blue). The cross shows the density peak of LAEs identified in Yamada et al. (2012).
}
\label{fig_finalimage}
\end{figure}

 We will describe the observation and the data analysis method in section \ref{sect:observation}. The sample selection by the photometric redshift (photo-z) as well as by the simple color cut will be described in section \ref{sample}. The properties of all the selected objects and $K_s$-band counterparts of LABs will be explained in section \ref{results}. In section \ref{discussion} we will discuss our results in terms of the mass assembly history in the protocluster. 

We use the cosmological parameter values $\Omega _{\rm M}=0.3$, $\Omega _{\Lambda}=0.7$, and $H_0$=70 km s$^{-1}$ Mpc $^{-1}$ throughout this paper. All the magnitude values are in {\it AB} system \citep{key-oke, key-fukugita}, unless explicitly noted. Conversions between the Vega and the AB systems are made using the following equations: $J=J_{\rm Vega}+0.95$, $H=H_{\rm Vega}+1.39$, $K=K_{\rm Vega}+1.85$.

\section{OBSERVATIONS AND DATA}
\label{sect:observation}

$J$, $H$, and $K_{\rm s}$-band images were obtained by using MOIRCS equipped with Subaru telescope, whose field of view is $4'\times 7'$. The observed fields are located in SSA22-Sb1 \citep{key-matsuda04}. The observations were carried out on June and August 2005, July 2006, September 2007, and May 2008. The summary of the observations is listed in Table \ref{tab_obs}. 

 We observed the six regions, which are referred as SSA22-M1 to M6. The schematic feature is shown in Fig.\ref{fig_finalimage}. The name of the observed field is also shown in the figure. The area of the three fields, M3, M4, and M5, are $4'\times 3.5'$, since the half of the FOV  was not available in 2007 due to the detector trouble. The early results for the SSA22-M1 field, corresponding to the southern part of SSA22a \citep{key-steidel98}, was published in \citet{key-uchimoto08}. The additional data taken for the SSA22-M1 fields also analyzed in this paper.

 The images are reduced in standard manner with the MCSRED software package \footnote{http://www.naoj.org/staff/ichi/MCSRED/mcsred\_e.html} (Tanaka I. et al.). \citet{key-uchimoto08} can be referred for the detailed description. The effective area of the final image is 111.8 arcmin$^2$. The typical image size (FWHM) ranges from $0''.4$ to $0''.6$, and the limiting magnitude ranges from 24 to 25 mag. The details are listed in Table \ref{tab_limitmag}.

 For the detection and photometry, we use SExtractor (version 2.3) \citep{key-bertin}. The objects which have more than 16 connected pixels above 1.5 $\sigma$ in surface brightness are selected. The SExtractor MAG\_AUTO value is adopted as the $K$-band pseudo total magnitude. We calibrated our NIR data to the UKIRT photometric system \citep{key-tokunaga}. 

\begin{figure*}[tbp] \epsscale{1.3}
\includegraphics[width=17cm]{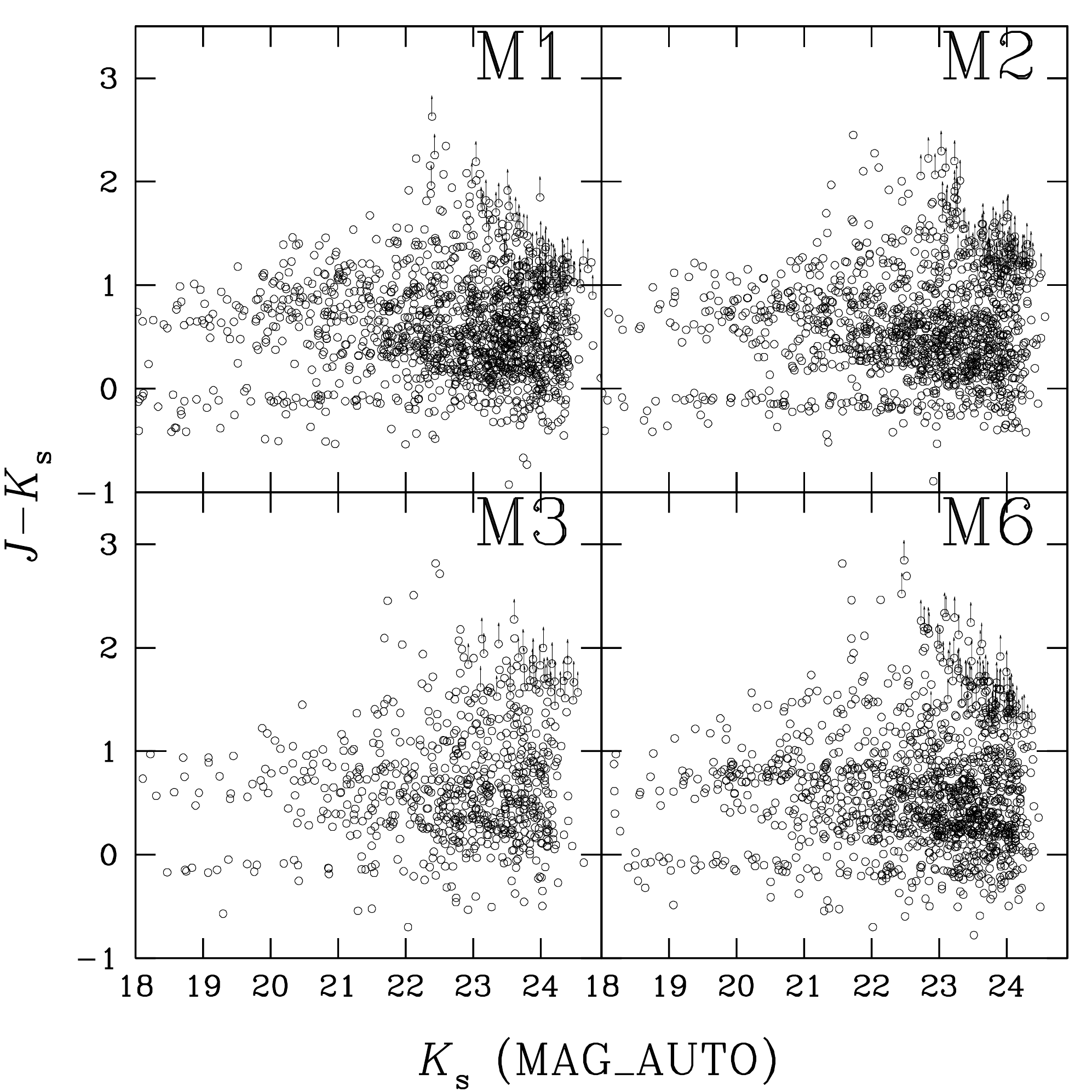}
\caption{
 $J-K$ vs. $K$ color-magnitude diagram of the $K$-selected objects in SSA-M1, M2, M3, and M6. The diameter of aperture photometry is 9 pixels ($1''.1$) is used for color measurement. The objects fainter than the $J$-band detection limit is show by the arrows. 
}
\label{fig_colmag}
\end{figure*} 
\begin{figure*}[tbp] \epsscale{} \center
\includegraphics[width=16cm]{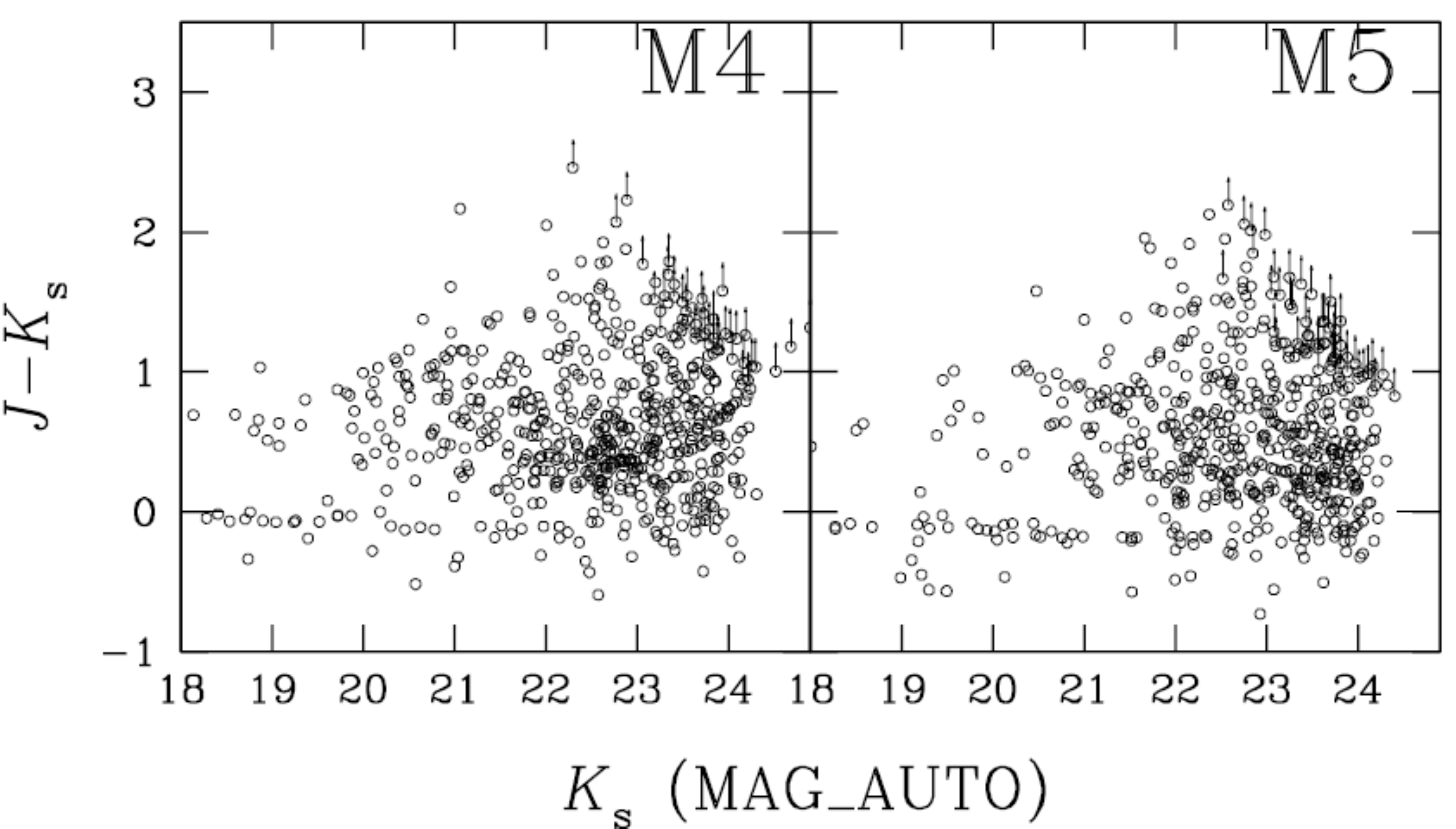}
\caption{
The same as Fig.\ref{fig_colmag}, but for
the $K$-selected objects in SSA22-M4 and M5.
The diameter of aperture photometry is 12 pixels ($1''.4$).
}
\label{fig_colmag2}
\end{figure*} 

 For the $J-K$ color measurement, all the images of the SSA22-M1, M2, M3, and M6 fields are smoothed to be matched with $ 0''.53$, the largest PSF of the $J$-band images in M1. On the other hand, FWHM in M4 and M5 are slightly larger as shown in Table \ref{tab_obs}. We therefore matched the PSFs in M4 and M5 to $0''.67$. The $J-K$ color was measured by a $1.1''$(9-pixels)-diameter aperture in M1, M2, M3 and M6, and by a $1.4''$(12-pixels)-diameter aperture in M4 and M5, and we applied  a small aperture correction less than 0.1 mag to the aperture magnitudes of the sources in M4 and M5. Fig.\ref{fig_colmag} shows the $J-K$ vs. $K$ color-magnitude diagrams for the four regions with better image quality, and Fig.\ref{fig_colmag2} shows that in the two regions with relatively poor image quality. 
 In SSA22-M1, M2, M3 and M6, the $J$-band images are deep enough to select the DRG sample ($J-K>1.4$) down to $K=24.0$. The completeness in the shallowest M4 field is still 90 \% at $K=23.8$.

 To estimate the photometric redshift, we also use the optical $BVRi'z'$-bands data obtained by \citet{key-matsuda04} with Subaru/Suprime-Cam as well as $u^*$-band data taken by CFHT/MegaCam (taken by P.I. L.Cowie). As the image size in optical and ultraviolet (UV) data is $1''.0$ in FWHM, we smoothed the $J$, $H$, and $K$ band images to be matched with the optical ones and measured the colors in $2''$ aperture for the purpose. 

 The Galactic extinction was corrected by adopting the average value at the field,  $E(B-V) = 0.08$ \citep{key-schlegel}.

\section{SAMPLE SELECTION}
\label{sample}

 We here select the potential protocluster members from the $K$-selected sample. Since $K$-band corresponds to rest-frame $V$-band at $z\sim 3$, the $K$-band luminosity of $z=3$ galaxies nearly reflect their stellar mass \citep{key-kajisawa09}. The nominal stellar mass limit at $K=24.0$ corresponds to $\sim 10^{9.5}$ M$_{\odot}$.

 We adopted the two kinds of selection criteria; one is the photometric redshift analysis by using $u^*BVRi'z'JHK$ data, and the other is the simple color-cut method. While full photometric redshift analysis is useful, it is still inevitable to have some scatters and a fraction of catastrophic failure if the photometric errors are ineligible and the limited model templates are used. As the photometric redshift needs the optical images with poorer image quality, it suffers from the relatively large photometric error due the larger photometric apertures (see above). A simple color selection such as DRGs thus works as a complementary method not only as they are simple and reproducible but also as the photometric error is relatively small especially for the fainter objects. 

\subsection{Photometric Redshift of the $K$-selected Sources}
\label{photoz}

 Using the multi-band photometric data, we obtained the photometric redshift of the $K$-selected sources using {\it hyperz} code \citep{key-bolzonella}. The SED fitting is performed with the redshift, spectral type, age, and dust extinction as the free parameters. The best fit SED is determined by minimizing $\chi^2$ value. The template spectra we used are those for the simple stellar population with the fixed (solar) metallicitiy from {\it GALAXEV} by \citet{key-bruzual}. 

\begin{figure}[tbp]
\epsscale{1.0}
\includegraphics[width=8cm]{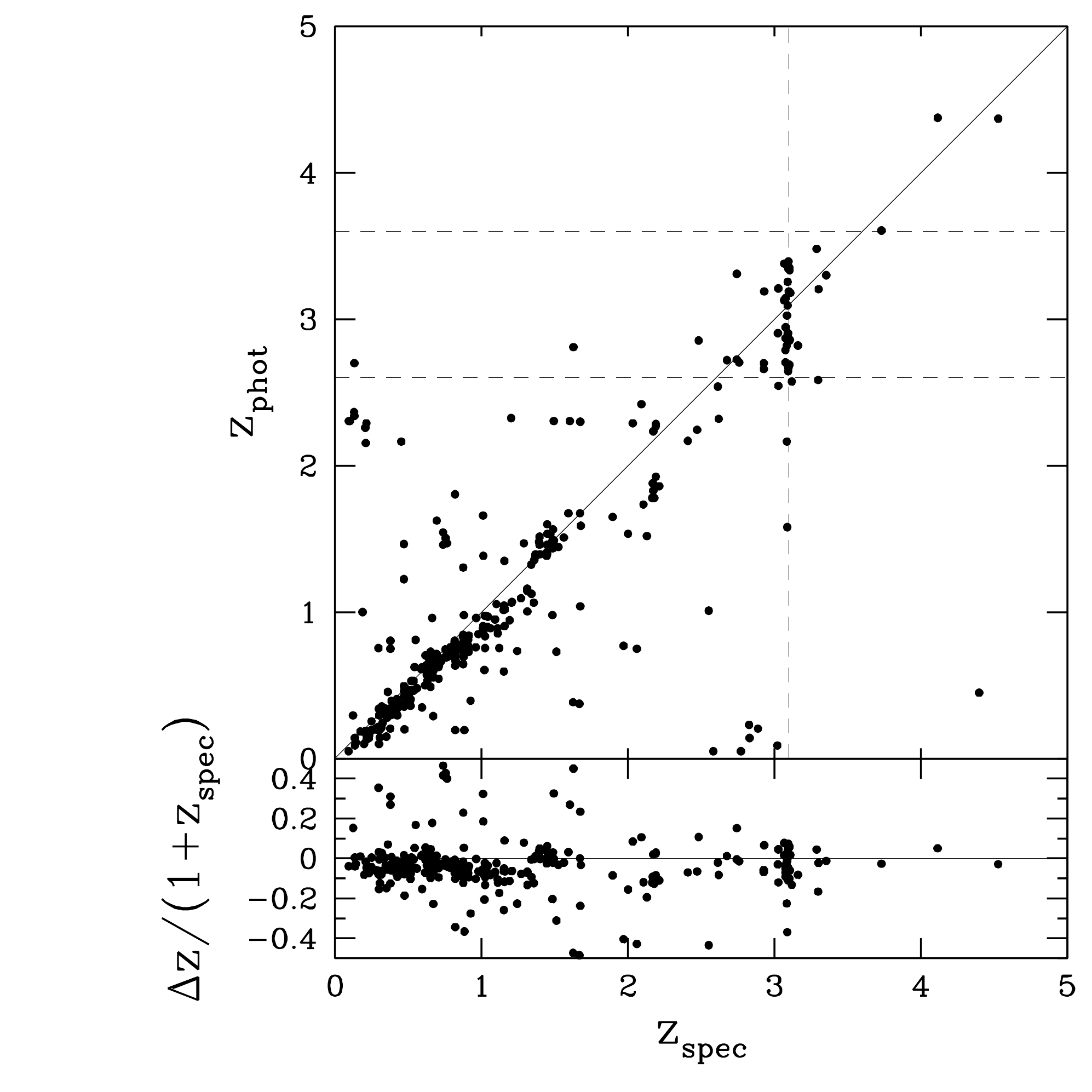}
\caption{
Comparison between photometric redshift and spectroscopic redshift 
for the $K$-selected objects in the SSA22-M1, M2, M3, M4, M5, and M6 fields.
The relative errors $(z_{\rm photo} - z_{\rm sp})/(1+z_{\rm sp})$ are 
indicated in the bottom panel.
}
\label{fig_speczphotoz}
\end{figure}

Fig.\ref{fig_speczphotoz} shows the photometric redshifts for the spectroscopic redshifts obtained by the previous literature and the NASA/IPAC Extragalactic Database (NED). The relative errors, $(z_{\rm photo} - z_{\rm sp})/(1+z_{\rm sp})$, are also shown in the bottom panel. The errors are similar if compared with those in the previous studies (e.g. Ichikawa et al. 2007). The uncertainty of $\Delta z \sim 0.5$ at $z\sim3$ remains since the Balmer/4000 \AA \hspace*{0.5mm}  break is shifted in the middle of $H$-band.

The median photometric redshift of 26 LBGs with $3.06<z_{\rm spec}<3.13$ is $z_{\rm phot}=3.11$ and the standard deviation is 0.08. 92 \% of the LBGs is in the range of $2.6 < z_{\rm phot} < 3.6$. We therefore pick up the objects with $2.6 < z_{\rm phot} < 3.6$ as the candidate galaxies at $z=3.1$. We finally selected 474 objects brighter than $K$=24.

\subsection{Selection of Distant Red Galaxies (DRGs)}
\label{drgselection}

 The red $J-K$ color of DRGs is due to the Balmer or 4000\AA\ break of galaxies at $z \gtrsim 2$, or to the dust extinction \citep{key-franx, key-reddy, key-forsterschreiber}. \citet{key-wuyts09} confirmed by spectroscopic observation that DRGs with $K_{\rm Vega}<22.5$ ($K_{\rm AB}<24.35$) are dominated by $z > 2$ galaxies. In the GOODS-North field, 86 \% of DRGs  are in the range of $2<z_{\rm phot}<4$ (Kajisawa et al. 2011), and the median photometric redshift is $z_{\rm phot}=2.5$.

\begin{figure}[tbp]
\epsscale{1.0}
\includegraphics[width=8cm]{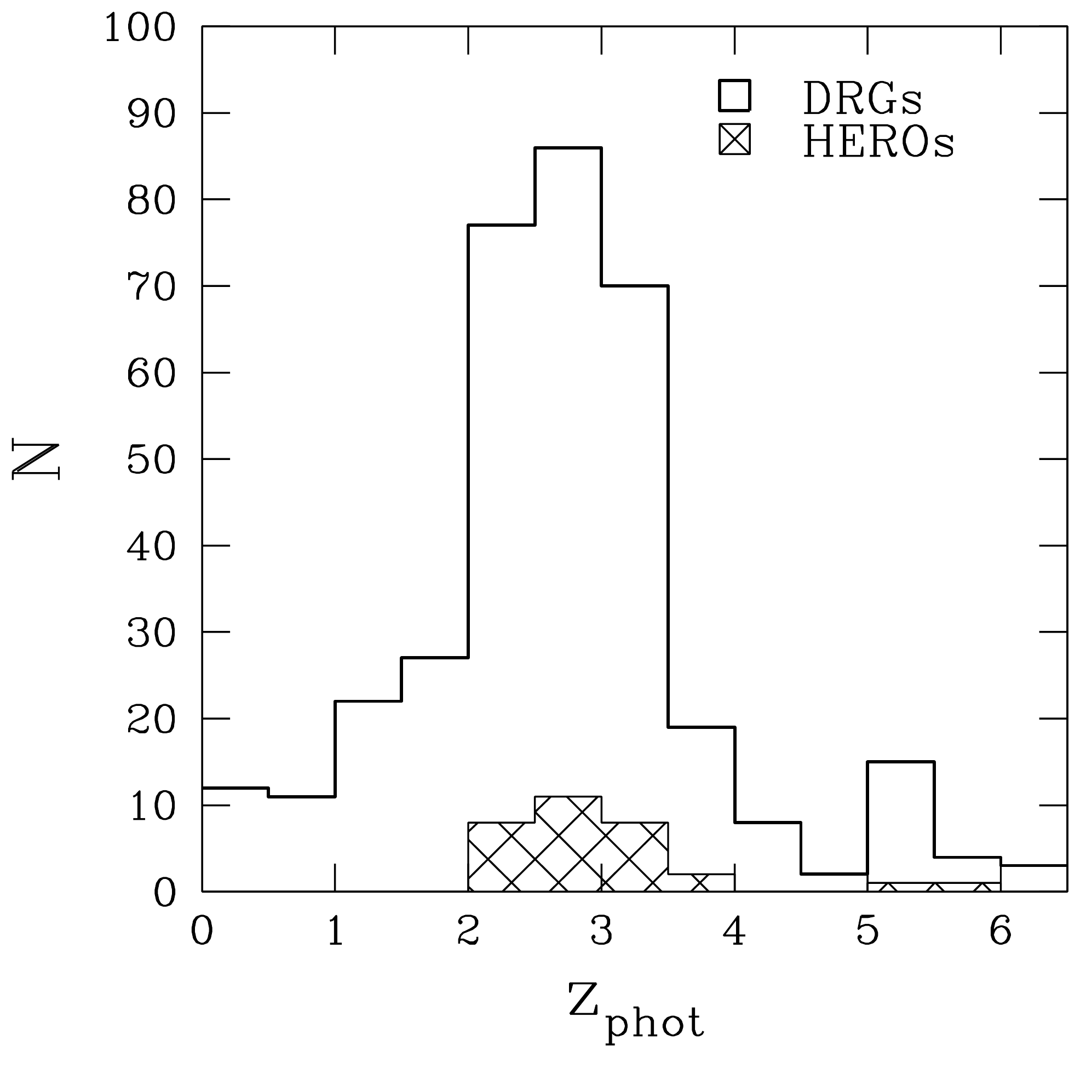}
\caption{
Histogram of the estimated photometric redshift of DRGs and HEROs
with $K<24.0$. The red histogram shows the numbers of DRGs and the hashed
histogram shows the numbers of HEROs.
}
\label{fig_photozdrg}
\end{figure}

 We used the same criteria as in \citet{key-kajisawa06}, which corresponds to $J-K_{AB}>1.4$ in our photometric system. 
 In our data, we sampled 356 DRGs with $K<24$ in the whole observed area. The distribution of the photometric redshifts for DRGs is shown in Fig.\ref{fig_photozdrg}.  We find that 70 \% of DRGs in the SSA22 area are located at $2 < z_{\rm phot} < 4$. The median photometric redshift is $z_{\rm phot}=2.6$.

 We compared the current sample of DRGs with the previously known sample of LBGs (Steidel et al. 2003). In the area of SSA22a, which has a substantial overlap with our MOIRCS fields, they listed in total of 171 LBGs including those without spectroscopic redshift (29 objects have the redshift between z=3.06 and 3.12). Of them, only the four objects (SSA22a-C54, SSA22a-M14, SSA22-MD39, and SSA22a-aug96C20) meets the $J-K$ color criteria. The $K$-band counterpart of SSA22a-M14a is, however, displaced from the optical source and may not be the same object or the same region in a galaxy. The spectroscopic redshift of SSA22a-aug96C20 is 1.357. No spectroscopic redshift is available for SSA22a-C54 and SSA22a-MD39.

 We also checked the optical colors of our sample of DRGs to see how many have the colors match the LBG criteria. Among the 356 objects with $K < 24$, 7 objects with $R<25.5$, magnitude limit for the spectroscopic sample of Steidel et al. (2003), and 23 objects with $R>25.5$ satisfies the LBG criteria adopted here,  $u^*-V > 2.9(V-R) + 0.37$, and $V-R < 0.5$, and $u^*-V > 1.4$. These numbers should be referred carefully as the color selection only properly works for those objects with high photometric accuracy ($>10\sigma$ in $V$ $\&$ $R$). The optical colors of those optically faint red objects are largely scattered due to the photometric errors. In conclusion, we found that the overlap between the current samples of DRG and LBG is very small.

\subsection{Selection of Hyper Extremely Red Objects (HEROs)}
\label{heroselection}

 We also sampled hyper extremely red objects (HEROs; Totani et al. 2001) with $J-K_{\rm AB}>2.1$ as a subset of DRGs. Totani et al.(2001) suggested that the extremely red color in $J-K$ is well explained by primordial elliptical galaxies reddened by dust and still in the starburst phase at  $z\sim3$. Indeed, in the GOODS-North field, all of the HEROs are in the range of $2<z_{\rm phot}<4$ (Kajisawa et al. 2011), and the median photometric redshift of HEROs is $z_{\rm phot}=3.0$.

 The number of HEROs in the observed field is 31. We find that 94\% (29/31) of HEROs in our field are classified as $2 < z_{\rm phot} < 4$. The median photometric redshift of HEROs is $z_{\rm phot}=2.7$.

 It should be noted that the result of the SED fitting for the HEROs tends to include relatively large uncertainty due to their extreme red color. 55\% of the HEROs are not detected in $J$-band at 2 $\sigma$ level (see Figs.\ref{fig_colmag} and \ref{fig_colmag2}). The number of the HEROs which are detected only in $K$-band is four while the rest of the objects are detected at least in the relatively deeper $R$-band image.

\section{RESULTS}
\label{results}

\subsection{The Distributions of the $K$-selected Objects in the SSA22 field}
\label{skydistribution}

\begin{figure*}[htbp]
\epsscale{1.2}
\includegraphics[width=9cm]{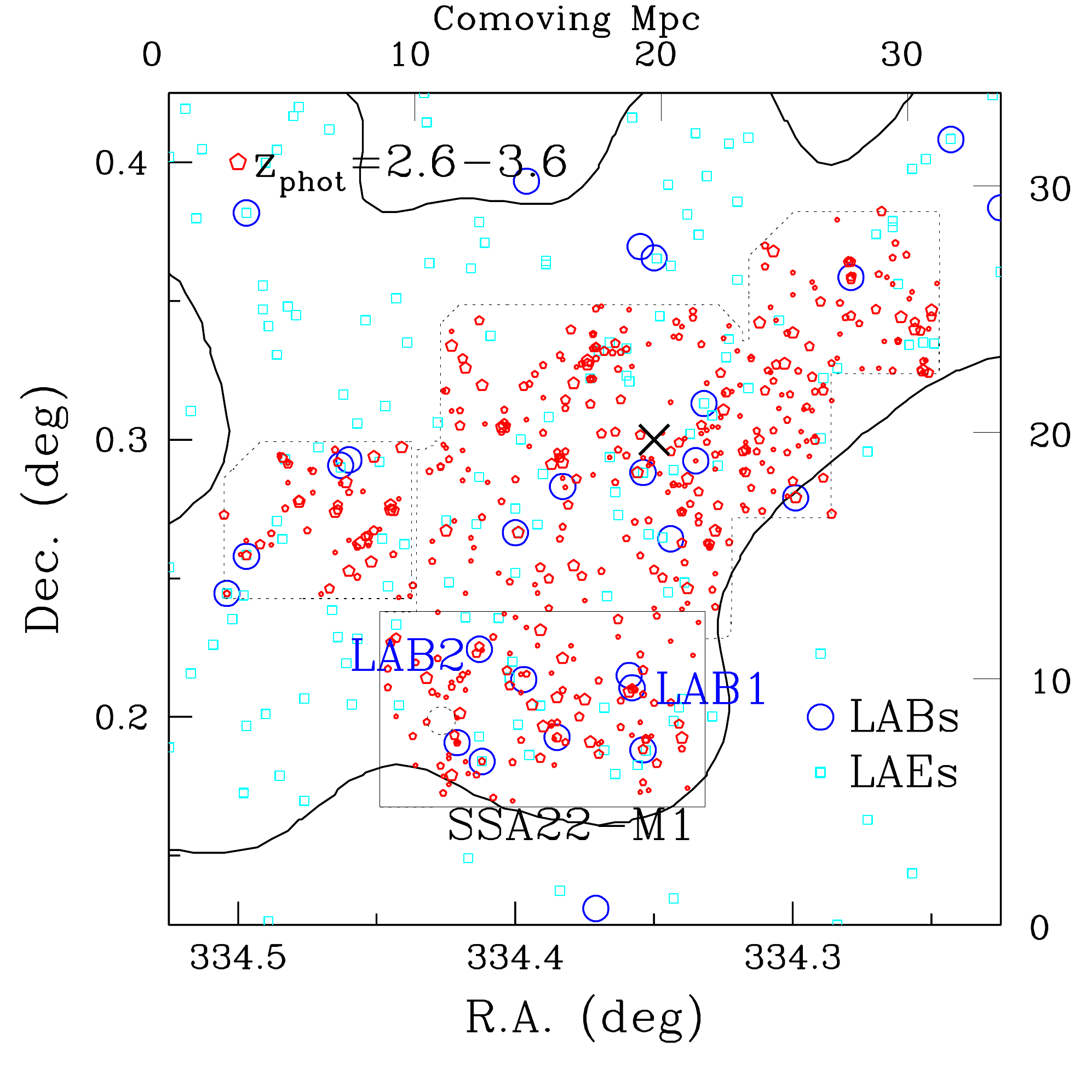} 
\includegraphics[width=9cm]{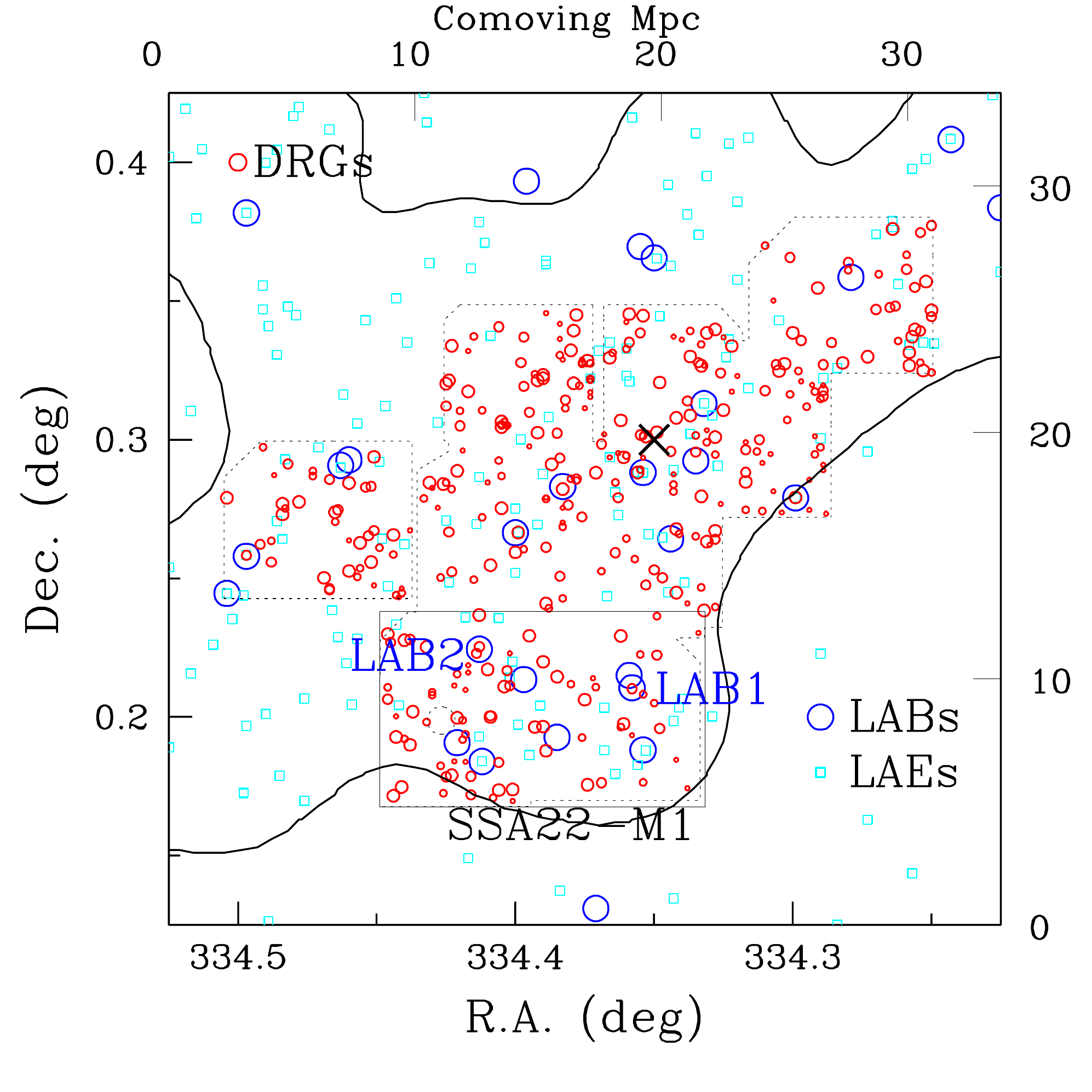} 
\caption{
(upper) The distribution of photo-z selected galaxies
($z_{\rm phot}=2.6-3.6$) (open pentagons, red).
(bottom) The distribution of DRGs (open circles, red).
LABs (Matsuda et al. 2004) and LAEs \citep{key-hayashino} are indicated with
large open circles (blue) and open squares (cyan), respectively.
The observed fields with MOIRCS are surrounded with the dotted lines.
The contour shows the high-density region of the LAEs \citep{key-hayashino}.
The cross is the density peak of LAEs identified in Yamada et al. (2012).
}
\label{fig_sky}
\end{figure*}

 The distributions of the $K$-selected objects with $z_{\rm phot}=2.6-3.6$ and DRGs on the sky are shown in Fig.\ref{fig_sky}, where LABs (Matsuda et al. 2004) and LAEs (Hayashino et al. 2004) are also plotted. The figure shows the samples with the magnitude limit of $K<24.0$. The dotted lines are the region observed with MOIRCS. We note that the noisy area at the center of the SSA22-M6 field is removed for the DRGs selection to avoid the spurious detection. The solid contours delineate the high-density region of the LAEs in Hayashino et al. (2004).

\begin{figure*}[tbp]
\epsscale{}
\includegraphics[width=18cm]{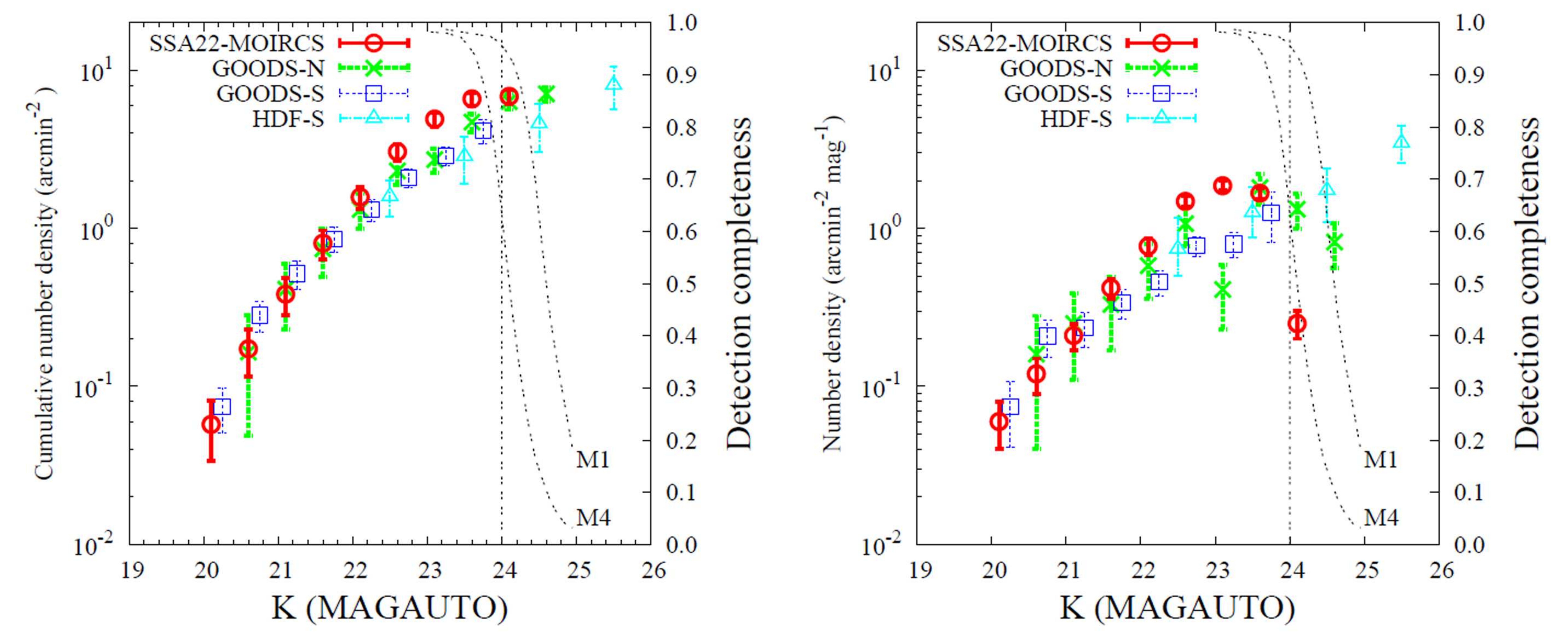}
\caption{
(left) Cumulative number counts of DRGs in SSA22-MOIRCS (M1, M2, M3, M4, M5, and M6) (open circles) in comparison with those for DRGs in GOODS-N (crosses), GOODS-S (open squares), and HDF-S (open triangles). The detection completeness in $K$-band in SSA22-M1 (deepest) and M4 (shallowest) is shown with the smaller dotted line. The Poisson errors are shown by the error bars.
(right) Differential number counts of DRGs in SSA22-MOIRCS. 
}
\label{fig_drgdensityall}
\end{figure*}

\begin{figure*}[tbp]
\epsscale{}
\includegraphics[width=18cm]{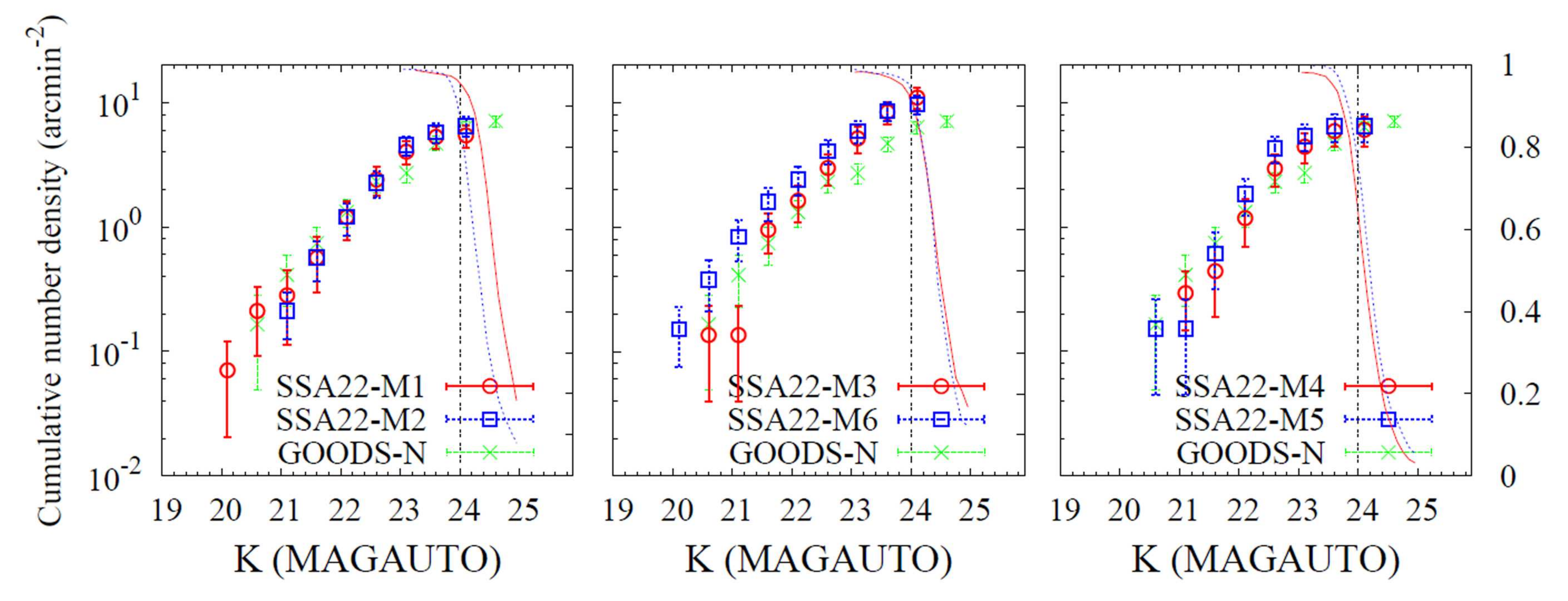}
\caption{
Cumulative number counts for DRGs in SSA22-M1, M2 (left), M3, M6 (middle), and M4, M5 (right)
 in comparison with those in GOODS-N (crosses). The number counts and the detection completeness in each field are shown with the same color. The error bars show the Poisson errors.
}
\label{fig_drgdensitycumula}
\end{figure*}

 In order to investigate whether there is an excess in number density of the galaxies in the SSA22 protocluster compared with the blank fields, we first investigate the surface number density of DRGs. The cumulative and differential number counts of DRGs in the whole field are shown in Fig.\ref{fig_drgdensityall}. The DRG counts show the excess at $K>23$ compared to those in all the three blank fields, namely the GOODS-North field (MOIRCS Deep Survey; Kajisawa et al. 2006, 2011), the GOODS-South field, and the HDF-South field \citep{key-grazian06}. The completeness fraction and the 90 \% limits in SSA22-M1 (deepest) and M4 (shallowest) are also plotted. The cumulative number density of DRGs in the entire observed field is $3.1$ arcmin${^2}$ down to $K=23.85$, which is $\sim$ 33 \% larger than that in GOODS-N.

 Fig.\ref{fig_drgdensitycumula} shows the DRG number counts in each field of view. The number density in GOODS-N and the 90\% completeness fraction are also plotted in the figure. The densest fields are SSA22-M3 and M6, which are overlapped with each other, followed by SSA22-M4, M2, M5, and M1. The number density of the DRGs in M6 is 2.1-1.8 times larger than that in GOODS-N at $K$=23-23.5. In addition, the excess of HEROs is found in this field (see section \ref{massall} and \ref{dusty}). Our observed fields cover the highest density peak of LAEs observed by \citet{key-matsuda11} and \citet{key-yamada11} in the 1.38 deg$^2$ survey area  (Fig.\ref{fig_finalimage}). The outstanding density peak of LAEs with the size of $\approx 3$ arcminutes lies in SSA22-M3 and M6, whose location is shown by the cross symbol in Figs.\ref{fig_finalimage} and\ref{fig_sky}. The fact that the largest overdensity of DRGs appears at around the highest density peak of LAEs suggests that the stellar mass distribution traces the structure of LAEs in the scale of several arcminutes, or a few Mpc at $z=3.1$.

 We also investigated the surface number density of photo-z selected objects and found that the counts of the objects at $2.6 < z_{\rm phot} < 3.6$ in the SSA22 fields are larger than those in the GOODS-N field (see Fig. 9 below). The direct comparison of the two samples should be discussed carefully, however, since the depth, or the photometric uncertainty of them is different.

 These results for the $K$-selected galaxies can be compared with the surface density of LBGs in the SSA22 fields. Steidel et al. (2003) found the average surface density of the color-selected LBG candidates with $R<25.5$ to be 1.72 arcmin$^{-2}$. While the surface density of the 11 fields observed with the Palomar 200-inch telescope ranges from 1.10 to 2.15 arcmin$^{-2}$, the density in SSA22a and SSA22b are 1.88 and 1.15 arcmin$^{-2}$, respectively. The higher contrast of the protocluster in the $K$-selected galaxies is due to the fact that the more stellar massive galaxies traced by the $K$-band sample have intrinsically much stronger clustering properties and may be easier to be dominated by a single large structure.

\subsection{$K$-band Counterparts of Ly$\alpha$ Blobs}
\label{lab}

\begin{figure}
\epsscale{}
\includegraphics[width=8cm]{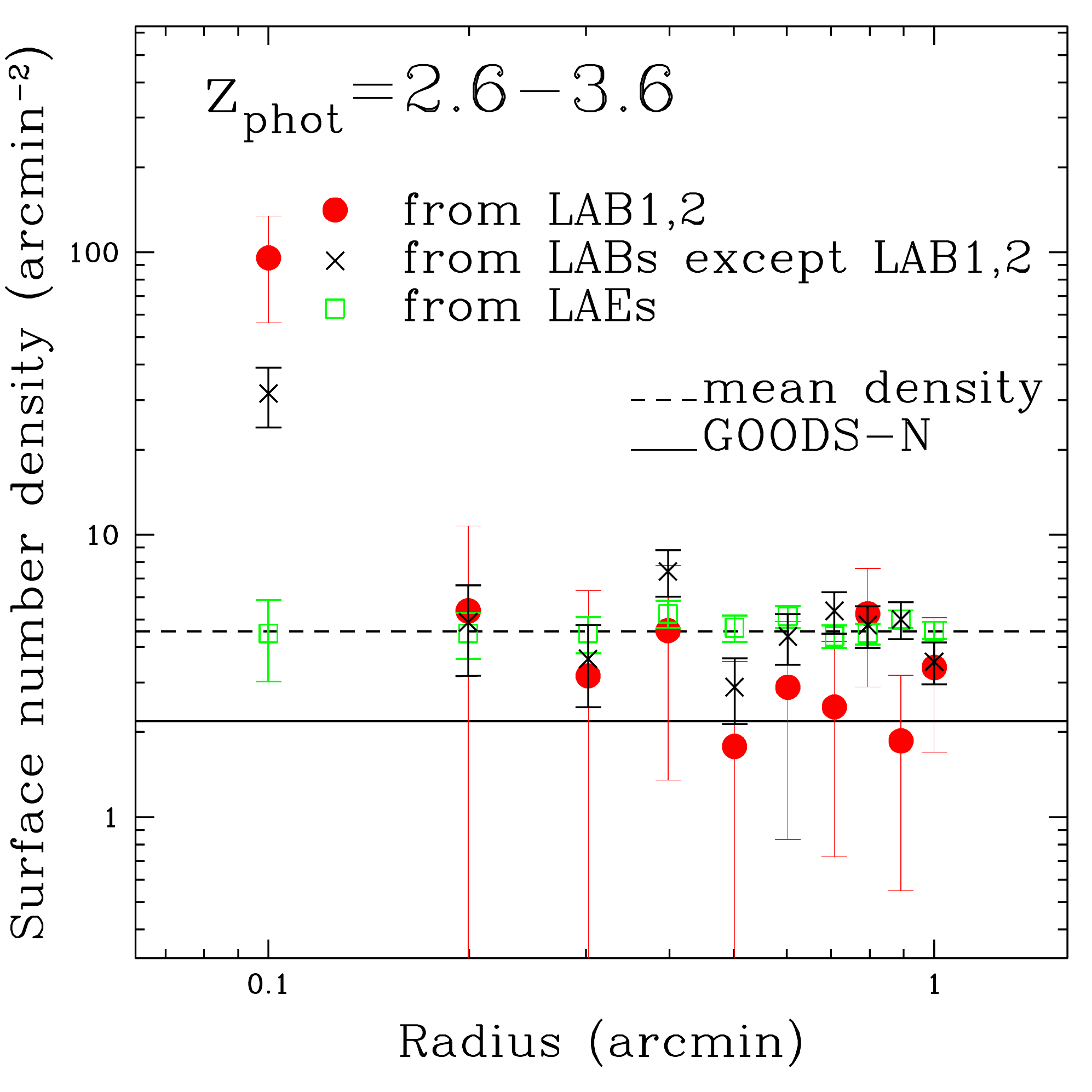}
\includegraphics[width=8cm]{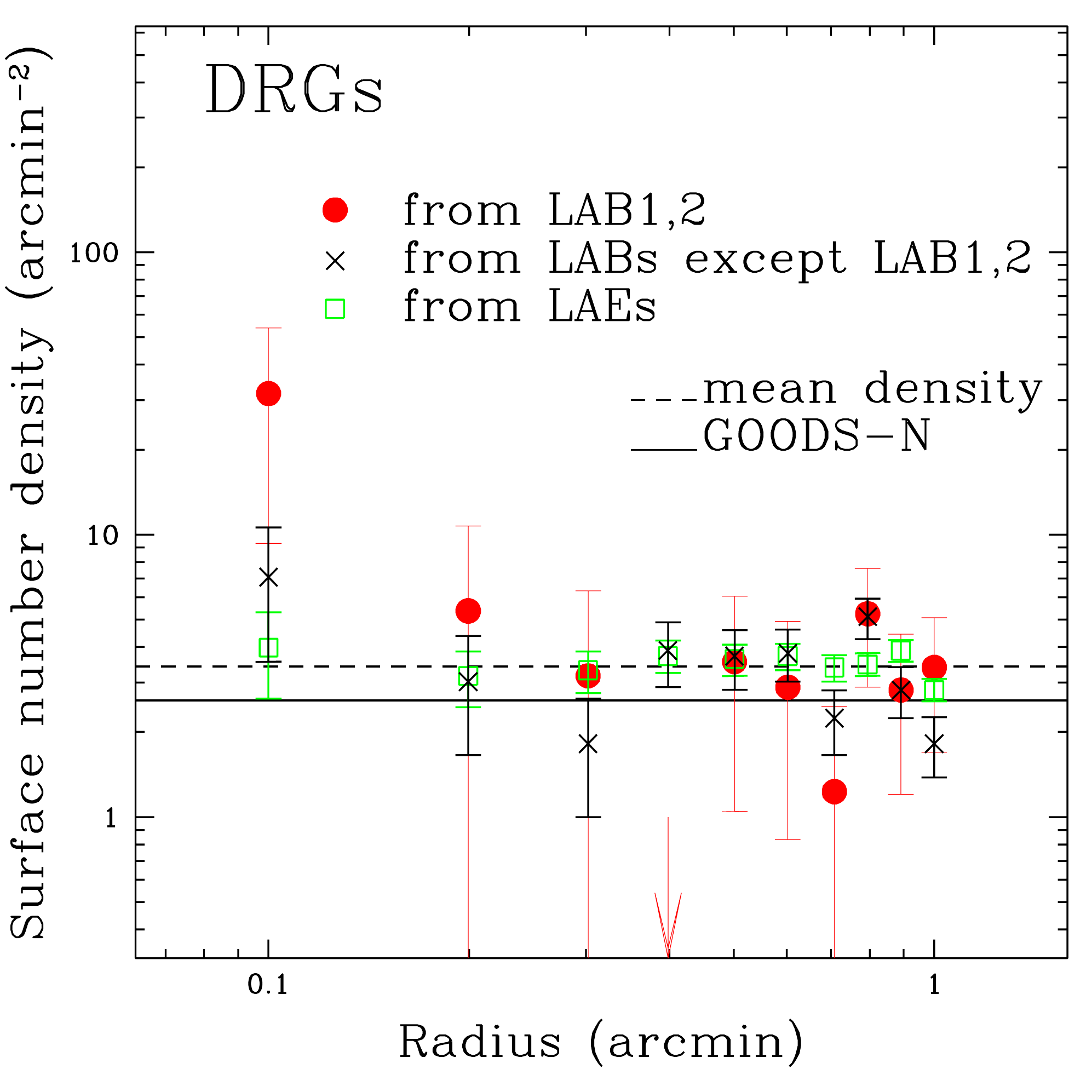}
\caption{
(upper) The surface number density of $K$-selected objects with $z_{\rm phot}=2.6-3.6$ in a circular ring of 0.1 arcmin width 
as a function of 
radial distance from LABs or LAEs.
The surface number densities 
from LAB1 and LAB2, LABs except LAB1 and LAB2, LAEs
are shown by filled circles (red), crosses (black), and open squares (green), respectively.
The mean densities of DRGs in SSA22 and GOODS-N are indicated with dotted and solid lines, respectively.
(bottom) The surface number density of DRGs in a circular ring of 0.1 arcmin width.
}
\label{fig_correlationdrg}
\end{figure}

 As seen in Fig.\ref{fig_sky}, photo-z selected objects and DRGs are frequently seen around the position of LABs. We investigate the spatial correlation between these $K$-selected objects and rest UV-selected galaxies (LABs and LAEs) in the field. Fig.\ref{fig_correlationdrg} shows the surface number densities of $K$-selected objects as a function of a radius from LABs and LAEs. As shown in the figure, we found that the $K$-selected objects are correlated with LABs within a radius of $6''$, which corresponds to 50 kpc scale at $z=3.1$. This implies that these objects are likely to associate with LABs. On the other hand, no significant correlation between LABs and $K$-selected objects at the larger scale is detected.

 In contrast to LABs, LAEs are not correlated with $K$-selected objects in scale of less than one arcminute while the large-scale density peaks of LAEs and $K$-selected objects coincide (section \ref{skydistribution}). Most of the LAEs are not detected in the $K$-selected images to the limit except for some HEROs (see section \ref{hero}). The expected typical luminosity of LAEs in the SSA22 protocluster is  $K\simeq 26.3$, which is estimated from the stacked image of LAEs that are not detected in $K$-band individually. The result is consistent with that of LAEs in a field environment (Ono et al. 2010).

 We here describe the detailed near-infrared properties of LABs. The number of the LABs (Matsuda et al. 2004)  in our observed fields is 20, and the $K$-band counterparts are detected for 75 \% (15/20) of them. Figs.\ref{fig_lab1to14} and \ref{fig_lab16to31} show the images of the 15 LABs, which have plausible $K$-band counterparts. Each panel has a length of 23.$''$4 on a side, which corresponds to 180 kpc at $z=3.1$. Circles are photo-z selected objects and DRGs with $z_{\rm phot}=2.6-3.6$ are colored red. Red squares are DRGs with $z_{\rm phot}<2.6$ or $z_{\rm phot}>3.6$. The contours are isophotal area of Ly$\alpha$ emission with surface brightness of 28.0 mag arcmin$^{-2}$ obtained by the narrow-band observation in Matsuda et al. (2004).

 Since the expected average surface number of DRGs or photo-z selected objects is only $\sim 0.03$ in the median Ly$\alpha$ isophotal area of 23 arcsec$^2$, or 0.0012 arcsec$^{-2}$ in surface density, it is interesting that a significant number of $K$-selected objects are detected around the LABs in our field. Table \ref{tab_labNIR} shows $K$-band magnitude, $J-K$ color, $z_{\rm phot}$, and $z_{\rm spec}$ of the $K$-selected objects in Figs.\ref{fig_lab1to14} and \ref{fig_lab16to31}. Spectroscopic redshifts of 11 photo-z selected objects including 9 LBGs listed in Table \ref{tab_labNIR} have been confirmed to be $z_{\rm spec}=3.1$.

 On the other hand, we detect no $K$-band counterparts for 5 LABs, LAB9, 19, 25, 26, and 35, as shown in Fig.\ref{fig_labnoK}. Please note that we plot all photo-z selected objects and DRGs even below the nominal magnitude limit. Detailed description for each object is given below.

\subsubsection{LABs with $K$-band Counterparts}
\label{labK}

 We identified the $K$-band counterparts of the LABs, which are numbered in Figs.\ref{fig_lab1to14} and \ref{fig_lab16to31}. There are the two types of LABs among them, those with a single or multiple $K$-band counterparts. LAB3, 5, 8, 14, 20, 24, and 31 are associated with a single component. The other 8 LABs, LAB1, 2, 7, 11, 12, 16, 27, and 30, have the multiple components.

\vspace*{2mm}

{\bf LAB1} has six $K$-band components within the Ly$\alpha$ halo. \#4, \#5 and \#6 are newly detected in this deeper $K$-band image. While the best-fit SED of \#5 gives the photo-z outside the criteria $z_{\rm phot}=2.6-3.6$, it is still a possible counterpart at $z\simeq 3.1$ if the large photometric errors in NIR are taken into account. There is a DRG (\#2) near the center of the LAB, and one LBG at $z=3.109$ (\#3) in the southwest from the center. As suggested in U08, a deficit of Ly$\alpha$ emission is found at the position of \#1. No X-ray source is detected \citep{key-lehmer09}, and there is no indication of AGN at the moment. MIPS 24 $\mu$m emission is detected at the position of \#1 as well as the region in the east from the center. The spectroscopic redshift at the Ly$\alpha$ emission peak and at the position of \#2 agree to be 3.097 \citep{key-ohyama03}. Near \#4, the optical continuum object is detected in a high resolution HST image \citep{key-chapman04, key-bower04}.

{\bf LAB2} is associated with five $K$-band components. \#1 is a DRG and the brightest in $K$-band, but the photometric redshift is slightly lower than our current criteria. \#2 is a DRG and the second luminous $K$-selected object. There is a LBG at $z=3.091$ in vicinity (indicated with a blue circle), $0''.9$ apart from \#2. \#3, \#4 and \#5 are newly detected objects. An X-ray source is detected near the position of the LBG, suggesting that there exists an AGN. MIPS 24 $\mu$m source is not detected at the position of the $K$-band sources.

{\bf LAB3} is associated with a single DRG, while two photo-z selected objects are located in vicinity in the east. This LAB is detected in X-ray, suggesting an AGN activity.

{\bf LAB5} is also associated with a single DRG. This object is detected in X-ray, suggesting an AGN activity. The shape of Ly$\alpha$ emission is elongated, or filamentary.

{\bf LAB7} is associated with the three $K$-band components. Detection of a LBG at $z=3.098$ was reported in \citet{key-steidel03}, but they are separated in two objects in our high resolution $K$-band image (\#1 and \#3). \#2 is also a LBG at $z=3.093$.

{\bf LAB8}, which can be to be a part of LAB1, has a LBG at $z=3.094$. The $K$-band counterpart is newly detected.

{\bf LAB11} includes three photo-z selected objects within the Ly$\alpha$ halo. The spectroscopic redshift of the unresolved optical counterpart is 3.0748.

{\bf LAB12} has three photo-z selected objects. \#1 is a DRG, and a LBG is marginally separated in two objects (\#2 and \#3). An X-ray source is detected at the position of \#1. \#1 is one of the most massive galaxies associated with LABs. 

{\bf LAB14} is associated with a single DRG, which is detected in X-ray and MIPS 24 $\mu$m \citep{key-webb09}. One of the most massive galaxies associated with LABs. 

{\bf LAB16} has the two photo-z selected objects within the Ly$\alpha$ halo. \#2 is newly detected.

{\bf LAB20} is associated with a single photo-z selected object, which is a LBG at $z=3.118$. The deficit of Ly$\alpha$ emission is seen at the position of the $K$-band source.

{\bf LAB24} is associated with a $K$-band source. Unfortunately, the color is not available as it is located at the edge of our images.

{\bf LAB27} has the four photo-z selected sources. The shape of Ly$\alpha$ emission is filamentary.

{\bf LAB30} has the two photo-z selected sources. \#1 is a LBG at $z=3.086$. \#2 is newly detected.

{\bf LAB31} is associated with a single photo-z selected object, which is a LBG at $z=3.076$. The shape of Ly$\alpha$ emission seems round.

\vspace*{2mm}

 Except for LAB1, the most massive 4 LABs (LAB2, 3, 12, 14) are detected in X-ray, which suggests that AGN activity. In addition, the luminous seven LABs have DRGs in their vicinity, suggesting that the LABs are associated with dusty starburst galaxies.

 The detection of the multiple $K$-band components in LABs suggests that we are witnessing "on-going assembly" or "hierarchical multiple merging" events of massive galaxy formation \citep{key-white78, key-kauffmann93, key-cole94, key-meza03, key-naab07}. The detailed discussion is given in section \ref{labmass}.

\subsubsection{LABs with no $K$-Counterparts}
\label{LABnoK}

 Five LABs, LAB9, 19, 25, 26, and 35, which have no photo-z selected counterparts, are shown in Fig.\ref{fig_labnoK}. Those LABs have no luminous NIR and optical counterparts in Ly$\alpha$ halos, while LAB9, 19 and 35, have optical components within a halo which are classified as foregrounds in our analysis. In addition, LAB9 and LAB26 seem to be a close pair on the sky as shown in the right bottom panel in Fig.\ref{fig_labnoK}, while the object located between LAB9 and LAB26 is also supposed to be a foreground object in our analysis. Additionally, the five LABs are not detected either in X-ray \citep{key-geach09} or MIPS 24 $\mu$m \citep{key-webb09}. These LABs are supposed to be less massive objects with at least $\lesssim 10^{10}$ M$_{\odot}$. These objects can be predominantly excited by the cold gas accretion (\citet{key-goerdt10}). \citet{key-matsuda11} suggested that the filamentary morphology of the nebulae may be the sign of cold accretion, although not all of the five LABs have such filamentary structures.

 Among them, only LAB35 has plausible counterparts in vicinity, which are one LBG and two DRGs. The redshift of the LBG is spectroscopically confirmed as $z=3.027$, which is a bit outside of the protocluster system. In addition, one of the two DRGs, B in Fig.\ref{fig_labnoK}, is classified as a foreground in our photo-z. On the other hand, the other DRG, which is indicated as A in Fig.\ref{fig_labnoK}, has the photo-z in the range of $z_{\rm phot}=2.6-3.6$ and associated with an emission-line object. The question is whether the object A is related to LAB35 or not. The redshift of the object A  (either the mission line component or the continuum) has not been determined spectroscopically, while LAB35 is already known to be at $z=3.098$. Their projected separation is 66 kpc if they are at the same redshift. The Ly$\alpha$ emission of the two objects does not show any indication of their connection; the emission at the object A and LAB35 are clearly snapped from each other and there is no counterpart between them in any wavelength.

\begin{figure*}[htbp]
\epsscale{}
\includegraphics[width=18cm]{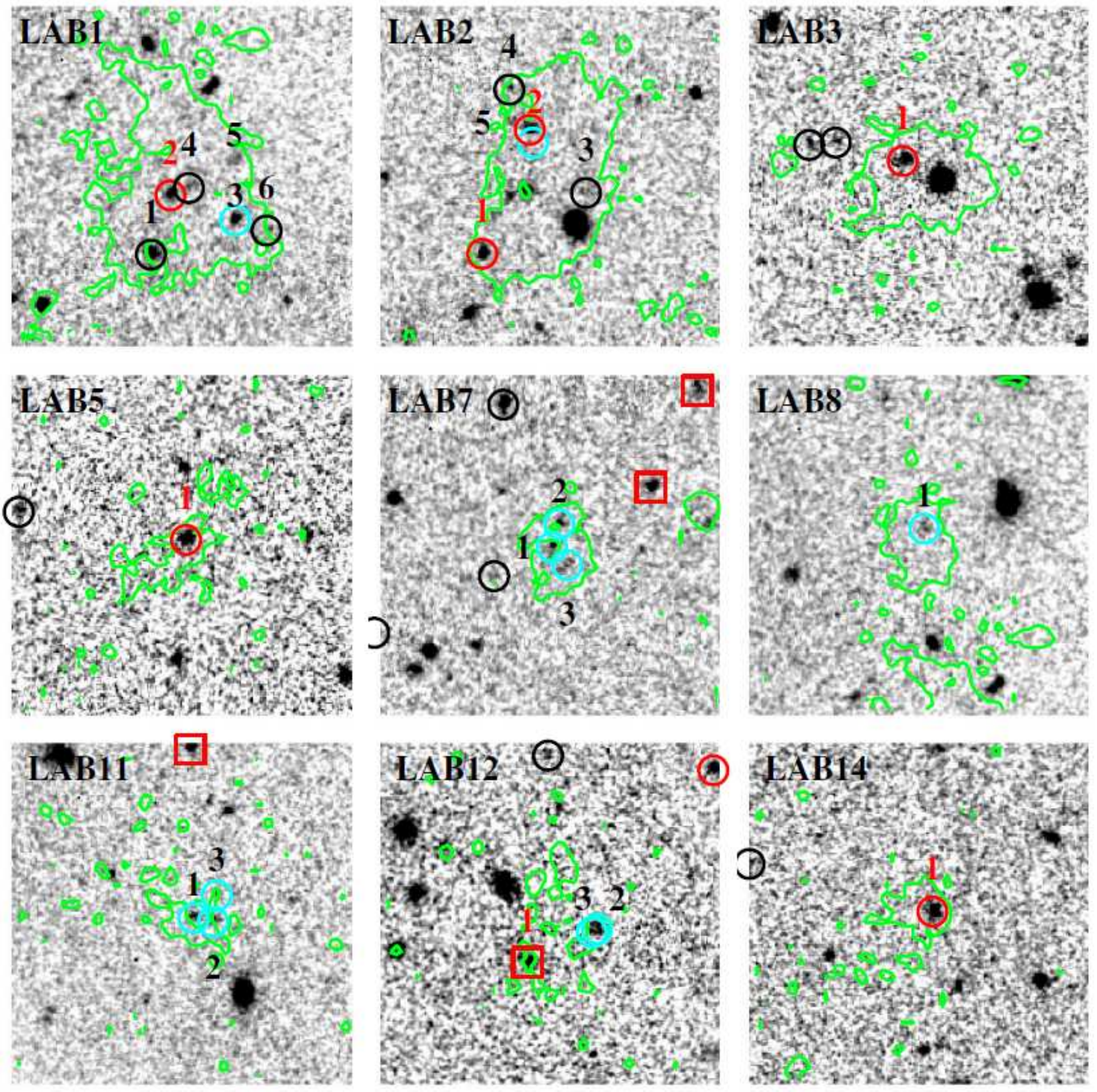}
\caption{$K$-band images around 15 LABs (Matsuda et al. 2004), which have 
$K$-counterpart candidates.
The size of each panel is $23''.4$, which corresponds to $180$ kpc
at $z=3.1$. 
The green contours are the isophotal levels of 28.0 mag arcsec$^{-2}$ in $NB497$.
The circles (black or red) indicate $K$-selected objects
with $z_{\rm phot}=2.6-3.6$.
The red circles are DRGs with $z_{\rm phot}=2.6-3.6$.
The red squares are DRGs at other redshift range 
($z_{\rm phot} < 2.6$ or $z_{\rm phot} > 3.6$).
The objects with numbers are adopted as $K$-band counterparts of the LABs.
The blue circle in LAB2 is centered on the position of LBG M14 in Steidel et al. (2003).
}
\label{fig_lab1to14}
\end{figure*}

\begin{figure*}[tbp]
\epsscale{}
\includegraphics[width=18cm]{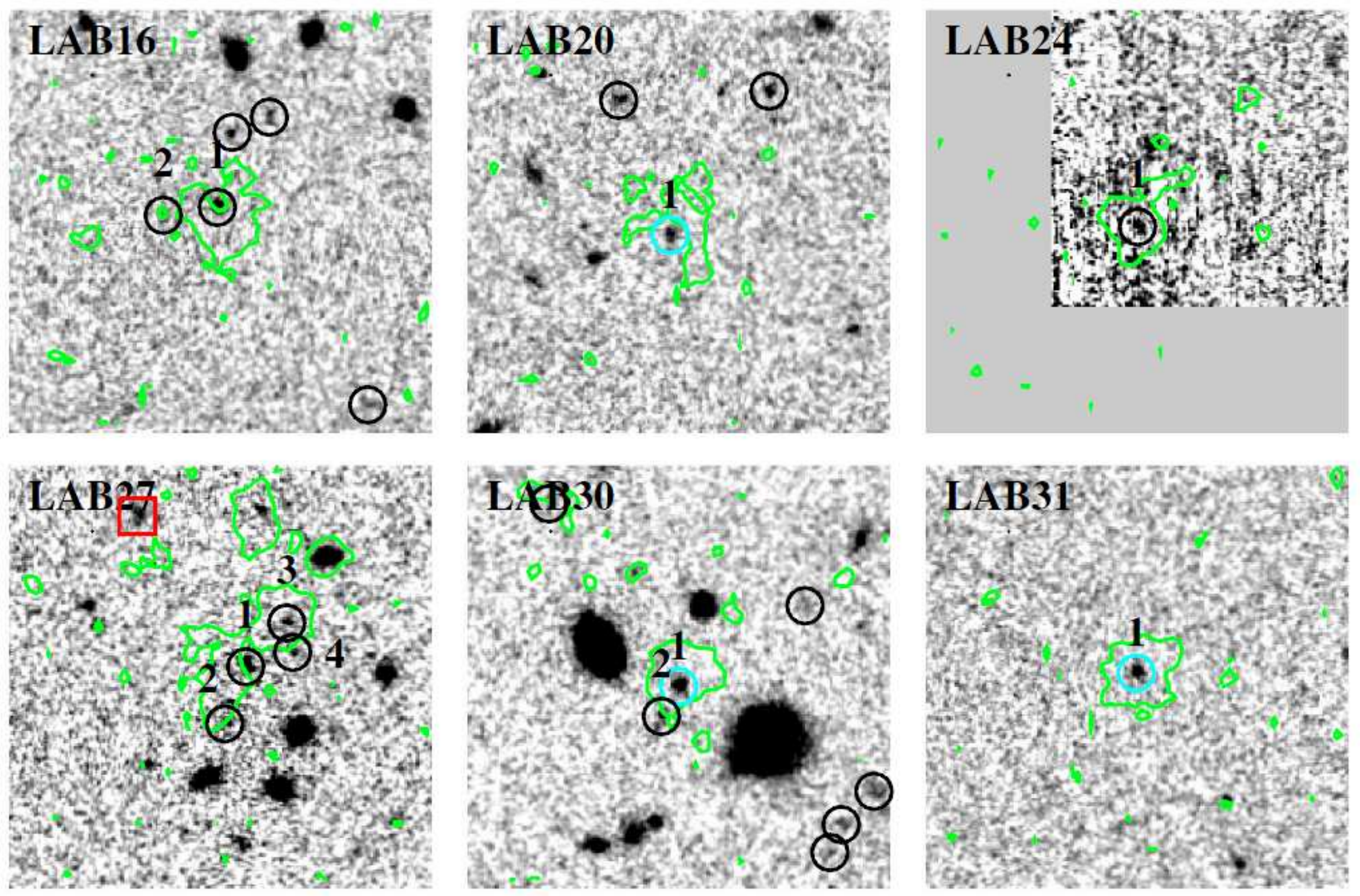}
\caption{ continued.
}
\label{fig_lab16to31}
\end{figure*}

\begin{figure*}[htbp]
\epsscale{}
\includegraphics[width=18cm]{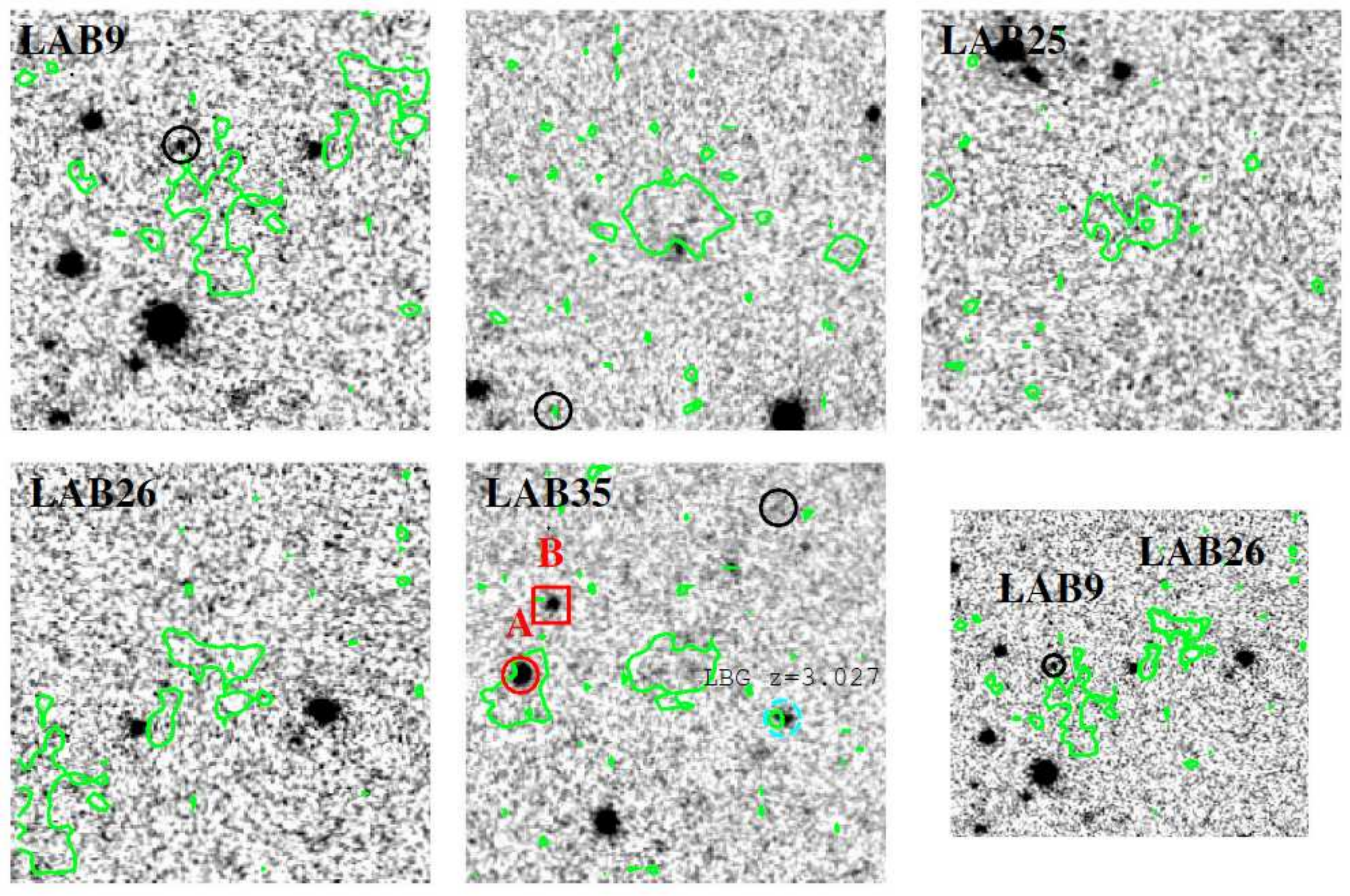}
\caption{$K$-band images around 5 LABs (Matsuda et al. 2004), which have  no $K$-counterpart candidates. The right bottom panel shows the close pair of LAB9 and LAB26. The symbols are the same as Fig. \ref{fig_lab1to14}. The LBG near LAB35 is identified as a foreground object at $z=3.027$ (see text).
}
\label{fig_labnoK}
\end{figure*}

\section{DISCUSSION}
\label{discussion}

\subsection{Massive galaxies in the SSA22 field}
\label{massall}

 In order to reveal whether massive galaxies have already been formed in the high redshift protocluster, we investigate the distribution of the stellar mass of the $K$-band selected objects in the SSA22-M1 to M6 fields. The stellar mass is derived from the SED fitting of the $BVRi'z'JHK$ photometric data. Spectral types, ages, and dust extinction are treated as free parameters. The redshifts are fixed at the photometric redshifts described in section \ref{photoz}, while the spectroscopic redshifts are used for LBGs if available. The redshifts of the $K$-selected objects in LABs described in section \ref{labK} are fixed at $z=3.1$. We used the template spectra obtained from GALAXEV \citep{key-bruzual} with star formation histories with $e$-folding time $\tau=$0, 1, 9 Gyr. The best-fit SED is determined from the minimum $\chi^2$ value. It should be noted that only the objects fitted with $\chi ^2 <1$ (85 \% of the sample) are used for the following discussion.

\begin{figure}[htbp]
\epsscale{}
\includegraphics[width=8cm]{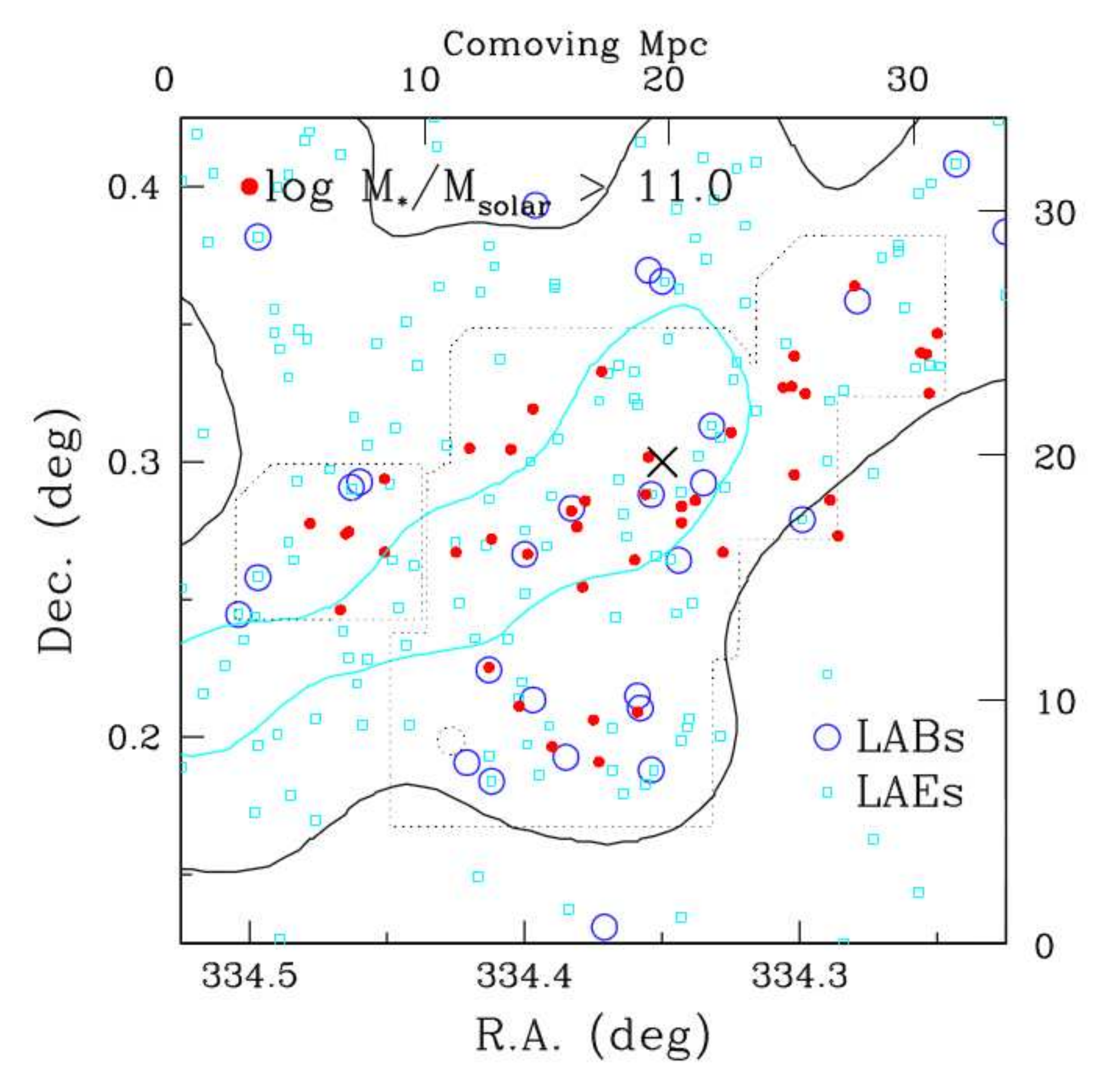}
\caption{
The sky distribution of  photo-z selected objects with $M_*>10^{11}$ M$_{\odot}$ (filled circles, red). LABs are indicated with large open circles (blue). The 1$\sigma$ and 2$\sigma$ density levels of LAEs are shown with the black and cyan contours. The cross shows the density peak of LAEs.
}
\label{fig_stellarmass_dist}
\end{figure}

 Fig.\ref{fig_stellarmass_dist} shows the sky distribution of the $K$-selected galaxies with $M_*>10^{11}$ M$_{\odot}$. These red massive galaxies are supposed to be the representative population in the SSA22 protocluster. While the number density of the galaxies with $M_*>10^{11}$ M$_{\odot}$in GOODS-N is 0.16$\pm$0.04 arcmin$^{-2}$, it was 0.37 arcmin$^{-2}$ in SSA22, 2.3 times larger. 

 In  Fig.\ref{fig_stellarmass_dist}, the surface number density of the massive galaxies is the highest near the peak of the LAE density (cross) although a little bit shifted to the south. There are 14 galaxies within 4$'$$\times$7$'$ ($\approx$ MOIRCS FoV) area, which gives the surface density of 0.5 arcmin$^{-2}$ or 3.1 times of the average of the GOODS-N field.
 
 On the other hand, we note that the number densities of DRGs and massive galaxies in SSA22-M1 are the smallest among the observed six fields. This is interesting as the SSA22-M1 field is the densest region of LABs and the two largest LABs (LAB1 and LAB2) locate in the area. The field is $\sim$10 Mpc (in comoving scale) away from the density peak of LAEs.

\begin{figure}[htbp]
\includegraphics[width=8cm]{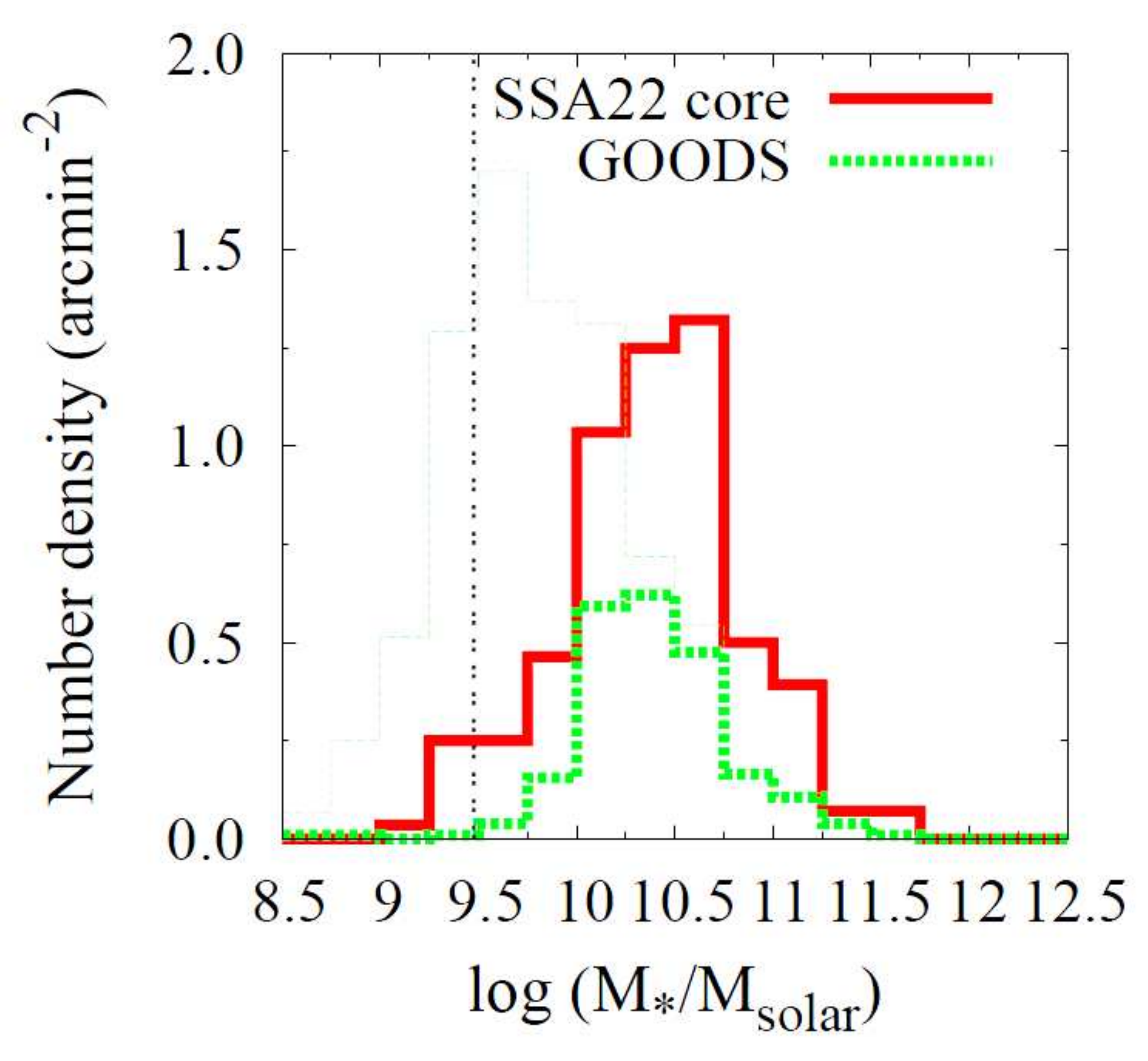}
\includegraphics[width=8cm]{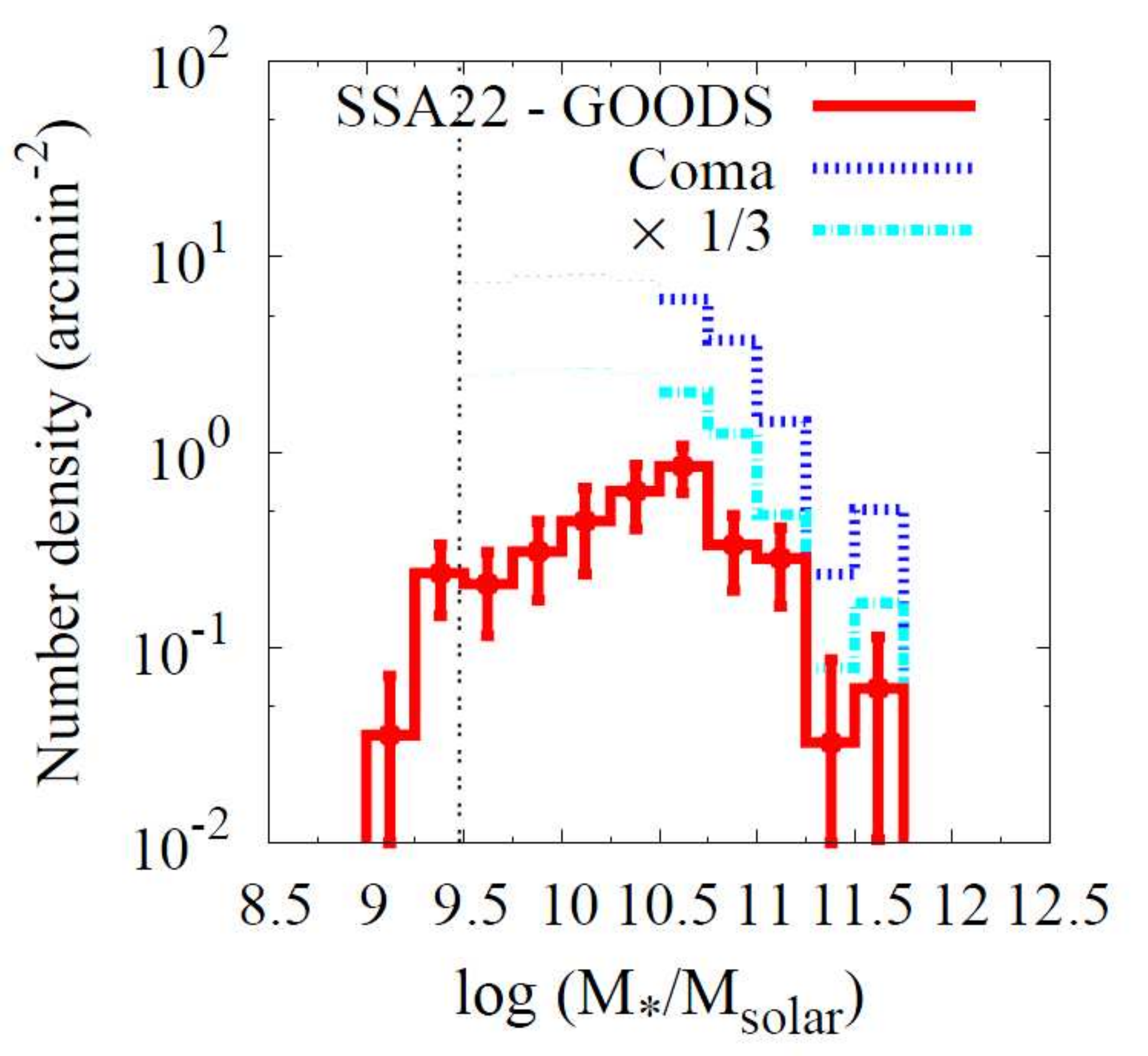}
\caption{(Upper) Histogram of stellar mass of photo-z selected objects ($z_{\rm phot}=2.6-3.6$) with $K<24.0$ in SSA22-M6 (the solid line, red). The vertical line is the nominal limit of the stellar mass at $K=24.0$. The dotted line (green) is the mass distribution in GOODS-N, and the thin line is that without the magnitude limit. (Bottom) The surface density difference between SSA22 and GOODS-N (the solid line, red). The stellar mass distribution in the Coma cluster is also indicated with the dotted blue line (see text). The same but divided by a factor of 3 is indicated with the cyan dot-dashed line.
}
\label{fig_stellarmass_histogram}
\end{figure}

 Massive galaxies in protoclusters at $z > 2$ have been identified in some other fields mostly around the radio galaxies. \citet{key-steidel05} identified a protocluster at $z=2.3$ in the HS 1700+643 field, which contains a significant number of old and massive galaxies. From their {\it spectroscopic} sample in the $\sim 70$ arcmin$^2$ field, there are at least three member galaxies with $> 10^{11}$ M$_{\odot}$. \citet{key-kodama2007} observed the field around the high-$z$ radio galaxies at $z=2-3$ with MOIRCS. They reported that there is an excess of galaxies with $>10^{11}$ M$_{\odot}$ on the bright end of the NIR color-magnitude sequence in the fields of MRC 1138$-$262 at $z=2.2$ and USS 1558$-$003 at $z=2.5$ while there are few massive galaxies on the red sequence in the protoclusters at $z=2.9$ and 3.1. For the MRC 1138$-$262 field, two of the massive red galaxies are identified as the protocluster members by MOIRCS spectroscopy \citep{key-doherty10}. Zirm et al. (2008) also investigated the HST NICMOS images of the field and found the significant overdensity of the red faint galaxies in the color-magnitude diagram. In addition, \citet{key-hatch09} derived the stellar mass of 19 galaxies in the Ly$\alpha$ halo of the radio galaxy MRC 1138$-$262, which is called as the Spiderweb galaxy (Miley et al. 2006), and found that there are no galaxies with $>10^{11}$ M$_{\odot}$ in the 0.3 arcmin$^2$ field except for the radio galaxy itself, which has the stellar mass of $10^{12}$ M$_{\odot}$. Compared to these previous studies, thanks to our wide field coverage, our results clearly identified the overdensity of massive ($>10^{11}$ M$_{\odot}$) red galaxies in a protocluster at $z=3.1$. Since this SSA22 protocluster is found as the largest high density region of star-forming galaxies known so far, the stellar mass assembly likely to be more progressed.

 Fig.\ref{fig_stellarmass_histogram} shows the histogram of the stellar mass distribution for the photo-z selected galaxies. The number density in the SSA22 peak  as well as that in GOODS-N  is shown with the solid and dotted lines, respectively. Both the samples are selected with the same selection criterion with $z_{\rm phot}=2.6-3.6$ and $K<24.0$. The turn off of the distribution at the less mass end (log M$_*$/M$_{\odot}$ $\lesssim$ 10.5) is due to the magnitude cutoff. The stellar mass distribution in GOODS-N without the cutoff is also indicated with the dotted thin line. It is found that the stellar mass distributions in both fields are not much different, while the number density in SSA22 is larger than that in GOODS-N. The median values of the stellar mass with $K<24.0$ in SSA22 and GOODS-N are 2.7 $\times 10^{10}$ M$_{\odot}$ and 2.2 $\times 10^{10}$ M$_{\odot}$, respectively.

 The excess in the surface number density of SSA22 to that of GOODS-N, which is regarded as the cluster component, is shown in the bottom panel of Fig.\ref{fig_stellarmass_histogram}. It might be interesting to compare this with the stellar mass distributions in nearby typical rich clusters. As an example, the stellar mass distribution for the Coma cluster is also shown in the figure. We focus on the massive end (log M$_*$/M$_{\odot}$ $\gtrsim$ 10.5) of the mass function because of the magnitude cutoff mentioned above. The stellar mass function of the Coma cluster galaxies is derived from the $H$-band luminosity function (LF) obtained by \citet{key-depropris98} (corresponds to $M_K^*=-24.0$, $\alpha=-0.78$ for the bright end) and the typical mass to light ratio of local elliptical galaxies ($M_*/L_K \simeq 0.6$). The number density is averaged in the observed area of $29'.2 \times 22'.5$ ($0.8 \times 0.6$ Mpc) in \citet{key-depropris98}, and converted to the physical density at $z=3.1$. It should be noted that the weight for the brightest galaxy in the Coma cluster in fitting LF seems low (Fig.2 in de Propris et al. 1998) and its contribution is not well reproduced by the obtained Schechter parameters. Therefore, we added it to the brightest bin manually. As shown in the figure, it is suggested that the equivalent of $\approx 10-15$\% of the galaxies with $>10^{10.5}$ M$_{\odot}$ in the Coma cluster is already seen in the SSA22 field. 
 In practice, however, the central region of the Coma cluster is supposed to be extended several times larger at $z\sim 3$. For example, \citet{key-suwa06} showed that a present-day cluster of galaxies with several Mpc can be extended to 20-40 Mpc in the comoving scale at $z=4-5$ in their simulation. By assuming the spherical model for the nonlinear collapse of the overdense region, the surface number density at the epoch of turn-around is four times lower than that of the collapsed object. If the protocluster does not reach the maximum expansion, it may be slightly larger. As for crude estimation, the stellar mass distribution in the Coma cluster divided by three is shown in the bottom panel of Fig.\ref{fig_stellarmass_histogram}. The figure suggests that the number of the $>10^{10.5}$ M$_{\odot}$ galaxies in SSA22 is equivalent to $\sim$ 30-40 \% of that in the Coma cluster.

\subsection{Stellar mass assembly in LABs}
\label{labmass}

 As shown in section \ref{lab}, most of the LABs are associated with the $K$-selected objects. The significant number excess around LAB that these galaxies are formed in LABs. It is also interesting to see a notable fraction of them consists of multiple $K$-selected components. The formation of early-type galaxies was generally understood by either ``Monolithic collapse'' (Eggen,Lynden-Bell, Sandage 1962) or ``(Major) Merger'' (Searle \& Zinn 1978) scenario. In the realistic cosmological scenario, however, early-type galaxies may have experienced hierarchical gaseous multiple merging in their early evolutionary phase, which may be observed as the combination of the extended gaseous halo and the multiple stellar components.

 Indeed, such phenomena is discussed using the cosmological numerical simulations for the formation of early-type galaxies \citep{key-meza03, key-naab07}. The scenario can be called as "on-going assembly" or "hierarchical multiple merging". The most outstanding examples observed in our observational data are the two giant objects, LAB1 and LAB2, which have 6 and 5 $K$-selected objects, some of which are DRGs, within their Ly$\alpha$ halo. As shown in the model simulations, these multiple components will finally merge into a single galaxy, which may evolve into a massive elliptical in the local universe.

 It is interesting to note that SSA22-M1 field where the LABs are the most populate is a few Mpc (in physical scale) separated from the center of the clustering of the $K$-selected galaxies or LAEs. The many (5/8) of the LABs in the SSA22-M1 field, LAB1, LAB2, LAB7, LAB16, and LAB30, in fact have the multiple $K$-band components. While the galaxy formation in the central region of the cluster proceed earlier statistically, we may see more dynamically young phase of massive galaxy formation in the region a little separated from the center.

\begin{figure}[tbp]
\includegraphics[width=8cm]{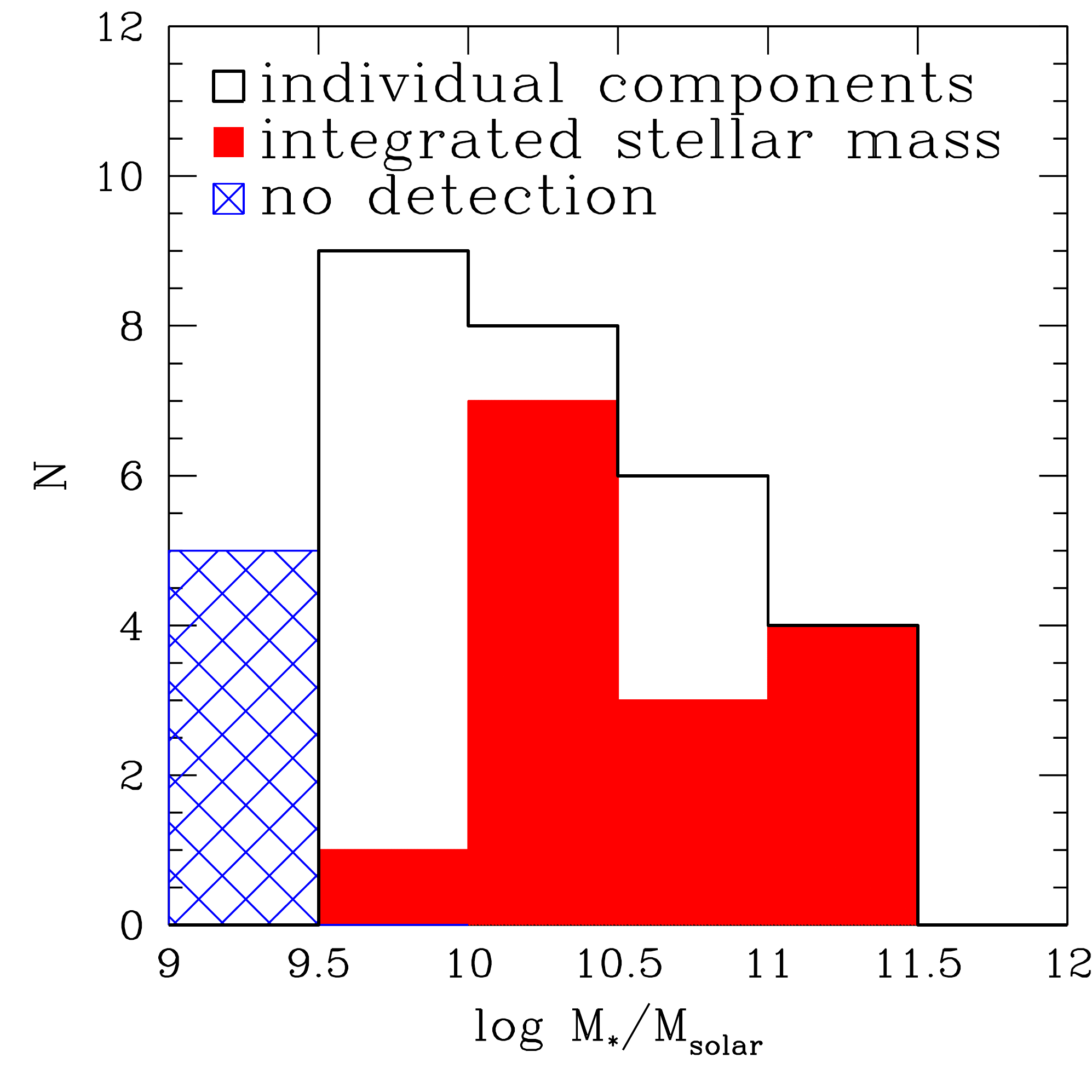}
\caption{
Histogram of stellar mass of photo-z selected objects in LABs.
 The stellar mass of each individual galaxy and 
the integrated stellar mass associated with the LABs
are indicated with filled and open histograms, respectively.
The hatched histogram indicates the number of LABs
which are not associated with photo-z selected objects.
}
\label{fig_hist_labmass}
\vspace*{3mm}
\includegraphics[width=8cm]{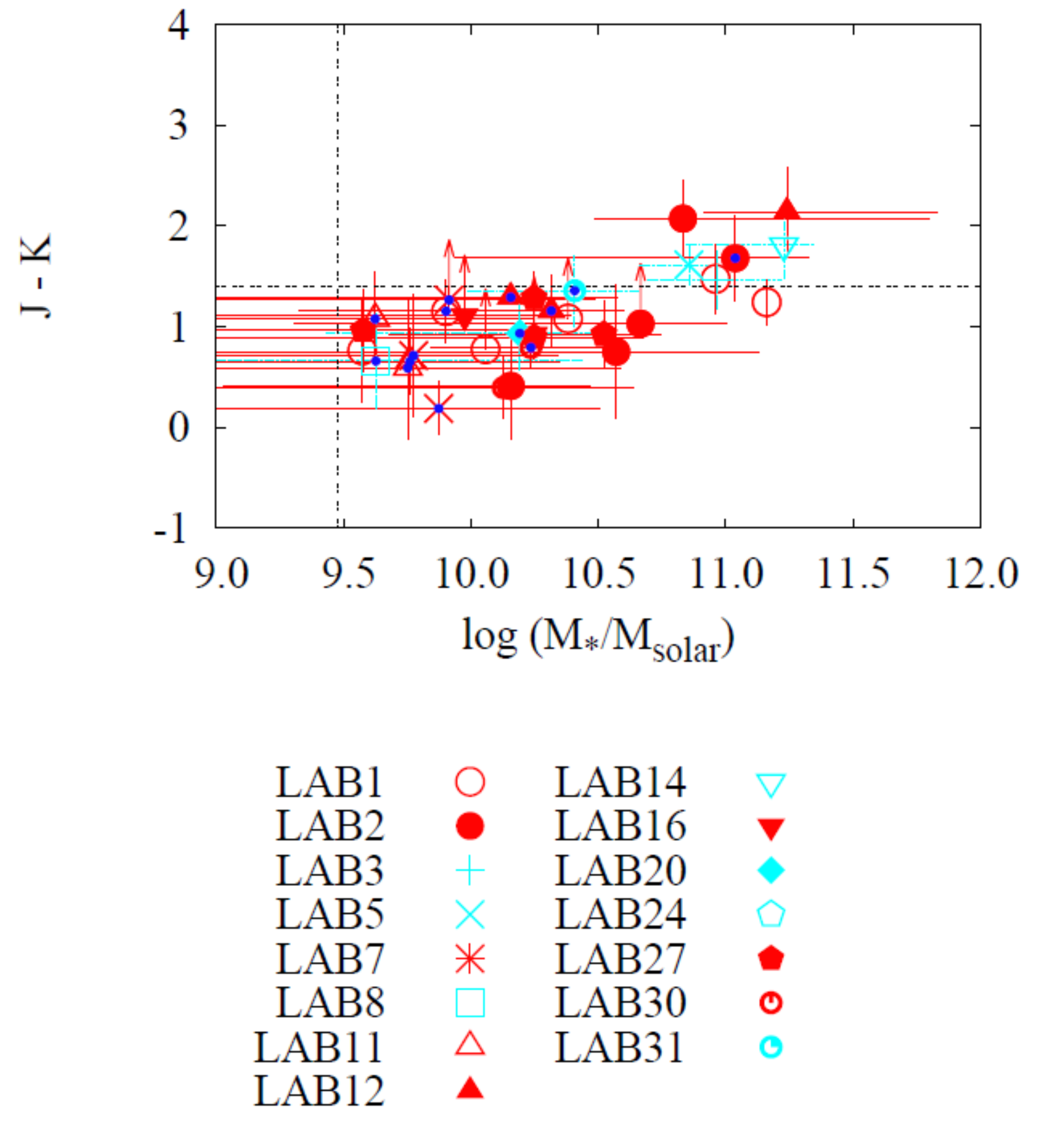}
\caption{
$J-K$ vs. stellar mass of each photo-z selected object 
in the Ly$\alpha$ halo.
The red symbols indicate that the LAB is associated with 
multiple $K$-band counterparts.
The cyan symbols indicate that the LAB has a single $K$-band counterpart.
The blue dots show that the objects are LBGs.
}
\label{fig_stellarmassvscolor}
\end{figure}

 We investigate the stellar mass of the photo-z selected objects associated with the LABs in order to investigate how much of stellar mass in these LABs have already been formed. The method is the same as that in the previous section, while the redshift is fixed to be $z=3.1$. The result is listed in Table \ref{tab_labSED}. We here assume that all the broad-band emission is due to the stellar light and ignored the contribution of the nebula emission lines, which will be constrained by the future spectroscopic observations.

 The stellar mass of the individual components of the LAB counterparts ranges from $3.8 \times 10^9$ M$_{\odot}$ to $1.7 \times 10^{11}$ M$_{\odot}$. Fig.\ref{fig_hist_labmass} also shows the distributions of the stellar mass of individual components and the integrated stellar mass associated with the LABs. Fig.\ref{fig_stellarmassvscolor} shows the color and the stellar mass of each individual component. LBGs are shown with the blue dots. As seen in the figure, low mass components in the LABs are dominated by LBGs. LAB7 and LAB11 have multiple LBGs with $< 10^{10}$ M$_{\odot}$, which are close to each other.

 With respect to the integrated total stellar mass in each LAB, 70 \% of the LABs (14/20) have  $> 10^{10}$ M$_{\odot}$, and the median value is $3.1 \times 10^{10}$ M$_{\odot}$. The 20\% of the LABs have the stellar mass $M_* > 10^{11}$ M$_{\odot}$. Typical LBGs in the general field have $M_* \simeq 2\times 10^{10}$ M$_{\odot}$ (Shapley et al. 2001), which is comparable or slightly smaller than the median stellar mass of the LABs of the current sample.

\begin{figure}
\includegraphics[width=8cm]{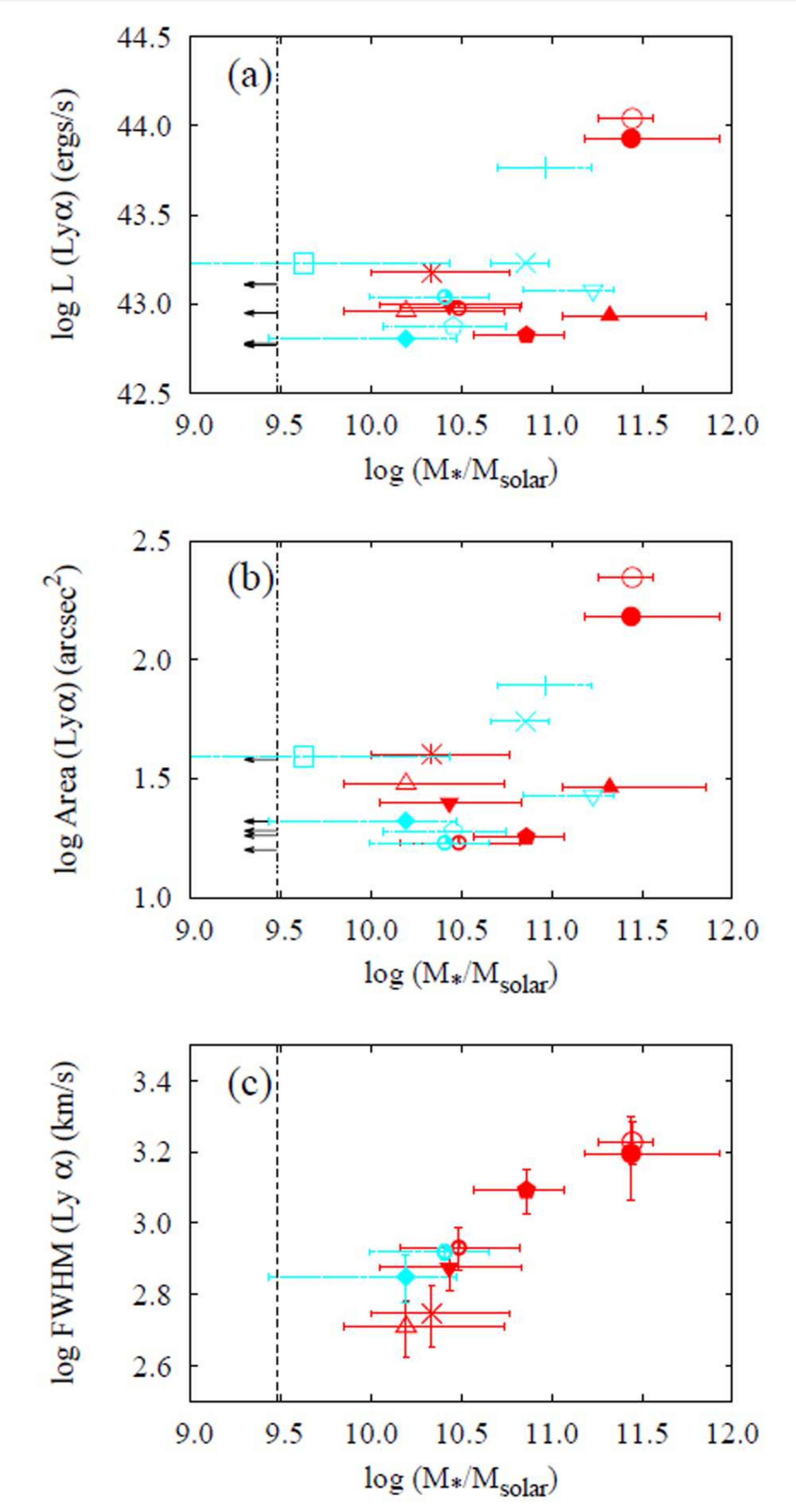}
\caption{
The relationship between the integrated stellar mass and the Ly$\alpha$ properties of LABs. 
(a) Stellar mass vs. Ly$\alpha$ luminosity, 
(b) Stellar mass vs. Isophotal area. 
(c) Stellar mass vs. FWHM of Ly$\alpha$ emission.
The symbols are the same as those  in Fig.\ref{fig_stellarmassvscolor}.
The multiple counterparts in each LAB are summed up in a symbol.
The vertical line is the nominal limit of the stellar mass 
at $K=24.0$.
}
\label{fig_lyastellarmass}
\end{figure}

 The total stellar mass in each LAB versus the properties of the Ly$\alpha$ emission are plotted in  Fig.\ref{fig_lyastellarmass}. The horizontal axes show the integrated stellar mass of photo-z selected objects associated with each LAB, i.e., the objects in Table \ref{tab_labSED}. The vertical axes in Fig.\ref{fig_lyastellarmass} (a), (b) and (c) show the luminosity, size, and velocity width of the LABs, respectively. Note that there is a correlation between the luminosity and the isopotal area of the Ly$\alpha$ emission of the LABs (Matsuda et al. 2004) and these are not totally independent. In Fig.\ref{fig_lyastellarmass} (a), or (b), we see that some massive galaxies with  $M_* \sim 10^{11}$ M$_{\odot}$ are associated with the brighter and larger LABs (LAB1, LAB2, LAB3). As also suggested in \citet{key-uchimoto08}, this implies that the origin of the Ly$\alpha$ emission in these LABs is related to the previous (or recent) star-formation activity represented by the accumulated stellar mass. On the other hand, LAB12 and 14 have large stellar mass of  $M_* \sim 10^{11}$ M$_{\odot}$ in spite of their low luminosity in Ly$\alpha$. Both LABs are associated with X-ray sources, and they are detected in 8 $\mu$m and 24 $\mu$m (Geach et al. 2009), suggesting that they suffer from dust obscuration. LAB14 is also detected as a SMG (Geach et al. 2005). In these cases, Ly$\alpha$ emission is supposed to be suppressed by the dust.

 In Fig.\ref{fig_lyastellarmass} (c), we can see that the total stellar mass and the velocity width of Ly$\alpha$ emission for 9 LABs obtained by Matsuda et al. (2006, also private communication) have the notable correlation. Here we should note that the velocity widths are estimated by assuming a single Gaussian profile, and the stellar mass estimation includes the uncertainty of photo-z. Despite such uncertainty, the correlation suggests not only that the photo-z selected objects are indeed associated with the LABs but also that the estimated velocity width represents their physical quantities.

  Although the correlation can be explained by some different scenarios \citep{key-matsuda06}, we first discuss about the case in which we assume that the Ly$\alpha$ gas clouds are gravitationally bounded. With the assumption, their dynamical mass can be evaluated from the velocity width and the radius which is represented by the extent of the Ly$\alpha$ emission. Then, our result suggests that the dynamical mass of LABs is correlated with their stellar mass. It is known that the dynamical mass of local spheroidal galaxies at $z<1$ correlates with the stellar mass  by a factor of $\sim 5 - 10$ \citep{key-drory04, key-bundy07}. On the other hand, the stellar mass of most of our sample are in the range of $10^{10.3} - 10^{11.5}$ M$_{\odot}$, and dynamical mass of LABs estimated by \citet{key-matsuda06} ranges over  $10^{11.7} - 10^{13.3}$ M$_{\odot}$. They are different by a factor of $\sim 50 -80$. As the ratio seems significantly larger than the local value, the stellar mass assembly may be still in progress in LABs.

 Of course, LABs may not be virialized, and the virial radius may not be equal to the size of the Ly$\alpha$ halo. 

 As an alternative case, if we assume that the Ly$\alpha$ emission is due to the outflow from galactic superwinds, the characteristic timescales of star formation can be estimated by their spatial extents and typical outflow speeds. \citet{key-matsuda06} evaluated the timescale of $\sim 10^{7-8}$ yr for the SSA22 LABs. Assuming the star formation rate of $10^3$ M$_{\odot}$ yr$^{-1}$ as inferred from the stacked sub-mm flux, the estimated amount of the {\it formed} stellar mass, $\sim 10^{10}$ M$_{\odot}$, is consistent with the timescale. In this case, it is suggested that the observed stellar components is formed from the star formation activities which cause the outflow.

\begin{figure*}[htbp]
\includegraphics[width=6cm]{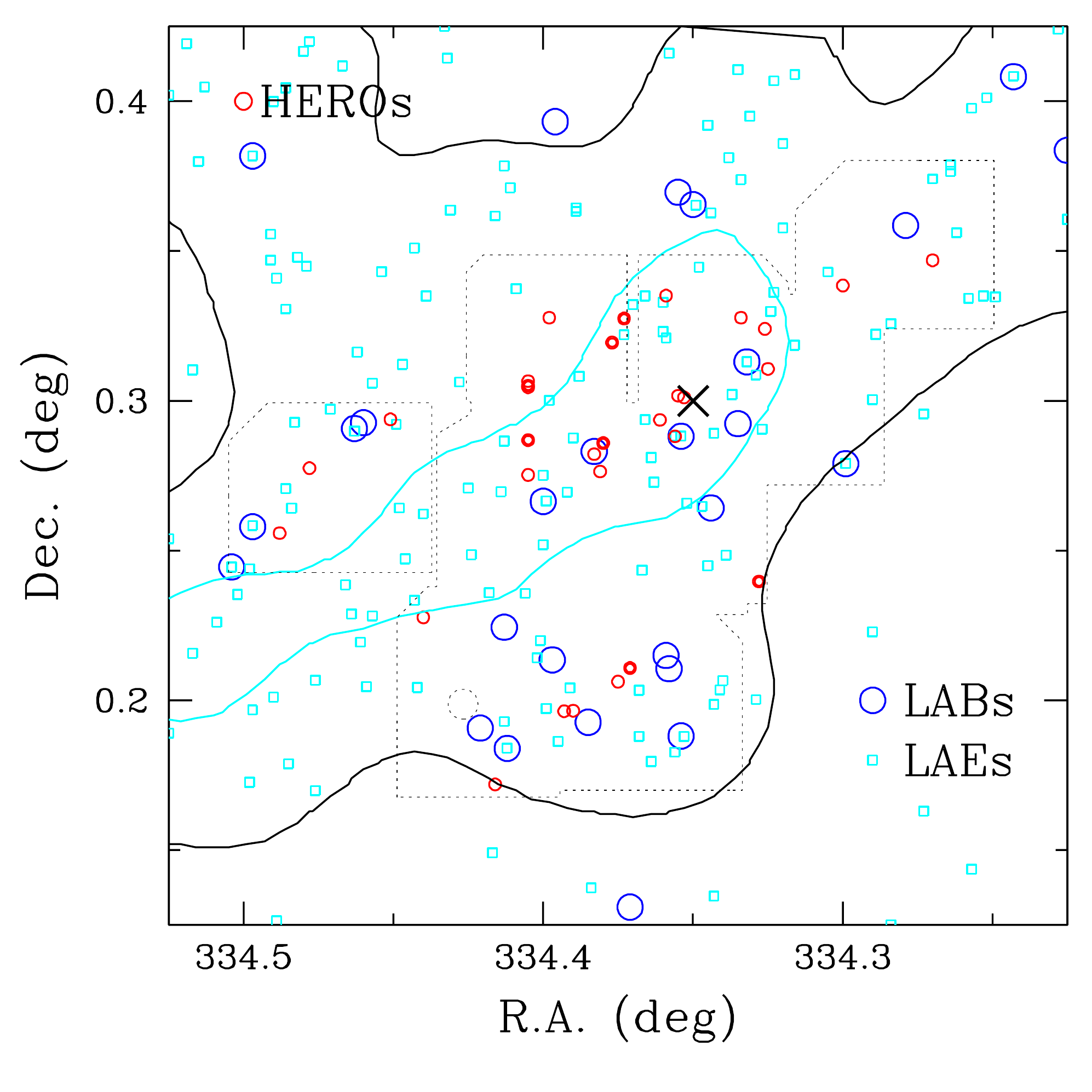}
\includegraphics[width=6cm]{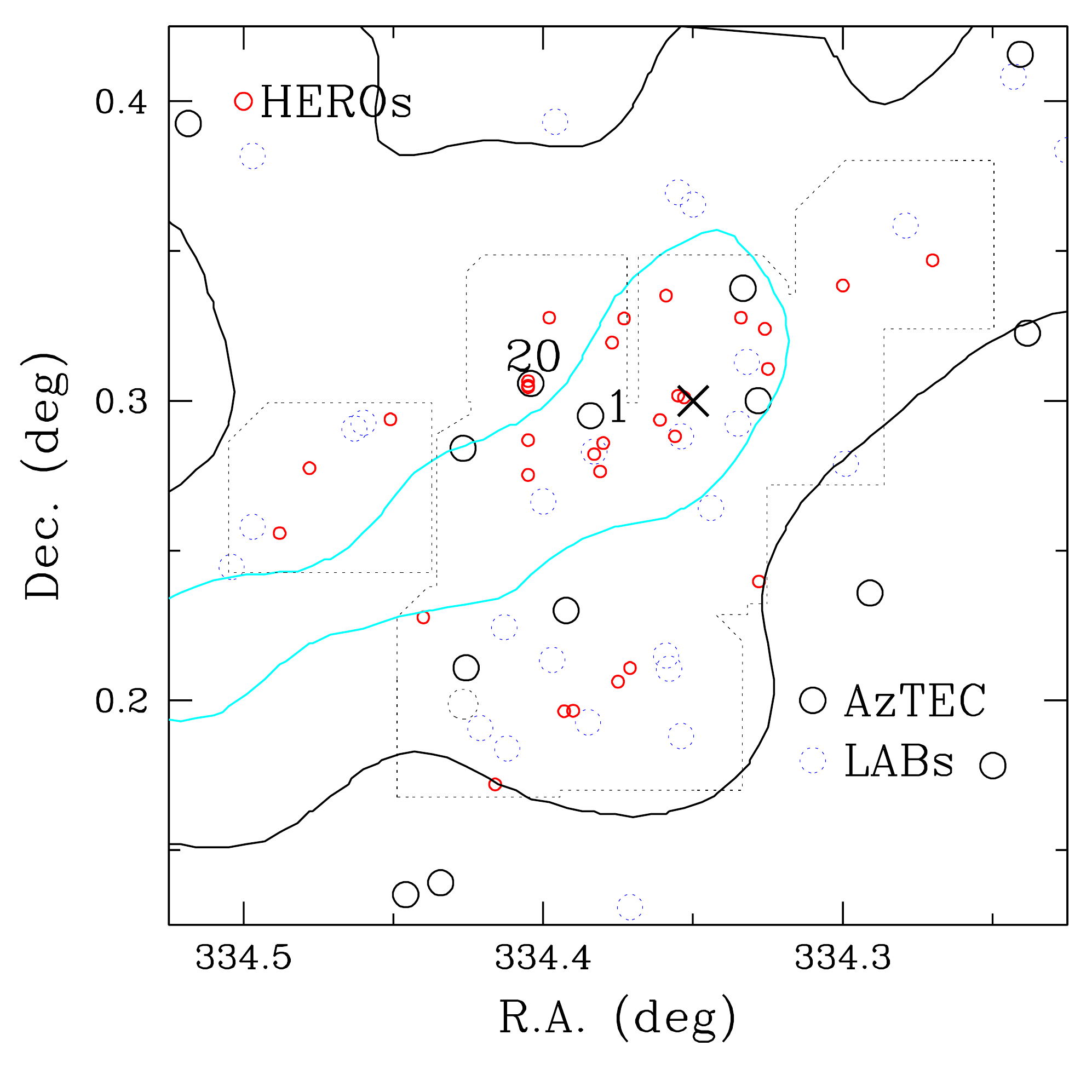}
\includegraphics[width=6cm]{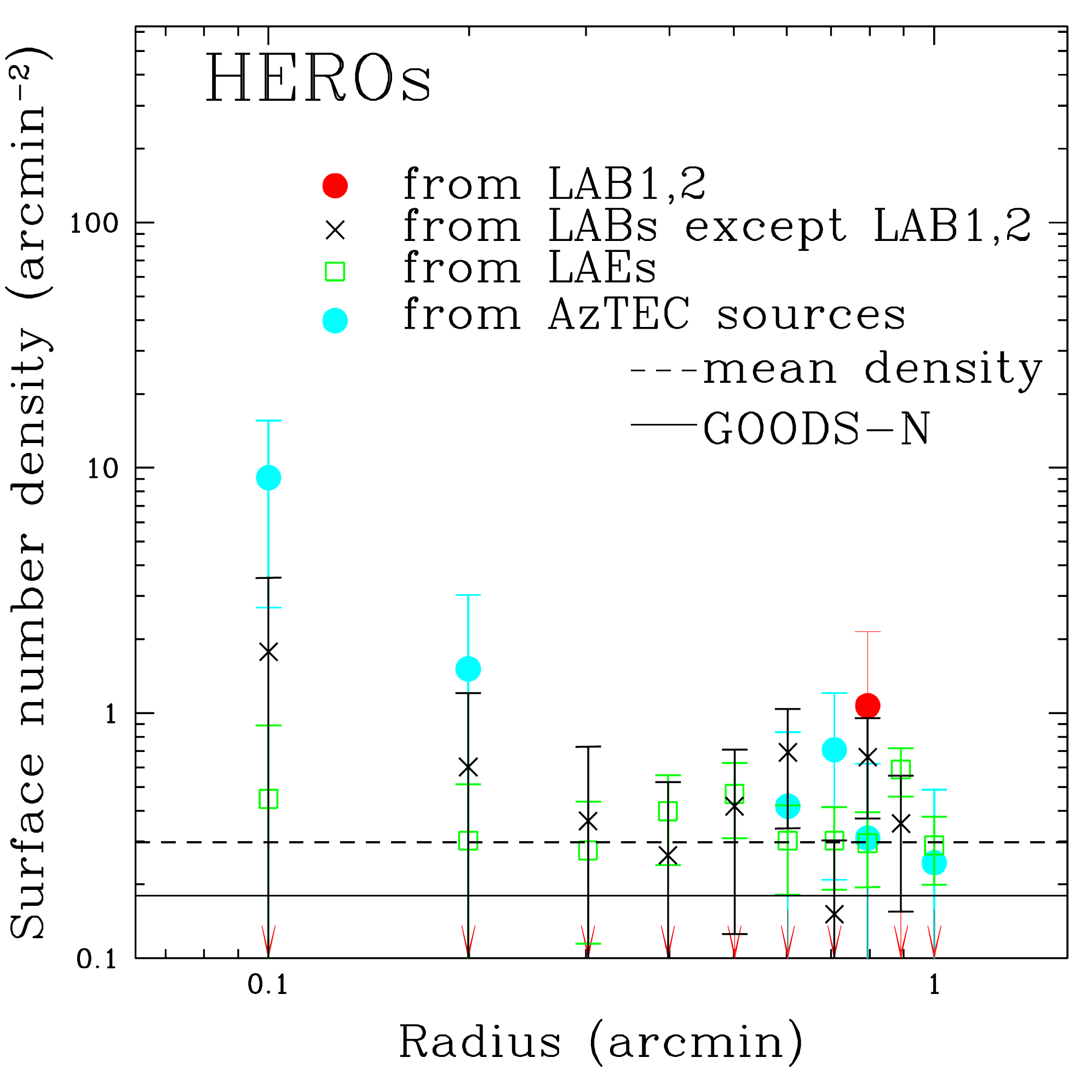}
\caption{ 
(left) Sky distribution of HEROs. HEROs are indicated with filled circles (red). 
LABs and LAEs are indicated with 
large open circles (blue) and open squares (cyan), respectively.
The contours show the high-density region of the LAEs \citep{key-hayashino}. The inner contour shows the 2 $\sigma$ level. The cross is the position of the density peak
of LAEs.
(middle) AzTEC sources are indicated with solid open circles (black).
(right) The surface density of HEROs in a circular ring of 0.1 arcmin width
as a function of 
radial distance from LABs, LAEs, and AzTEC sources.
}
\label{fig_skyheros}
\end{figure*}

\subsection{Dusty starbursts in the SSA22 field}
\label{dusty}

 $K$-band selection is also useful to detect the highly dust obscured starbursting galaxies. They are important population which dominates the stellar mass of the protocluster. We here focus on the dusty starbursts in our sample, namely, HEROs which are the objects with extremely red color of $J-K>2.1$ (see section \ref{heroselection}). We also discuss for the AzTEC sources detected in \citet{key-tamura09}, which are claimed to have the spatial correlation with Ly$\alpha$ emitters.

\subsubsection{Overdensity of HEROs in the SSA22 field}
\label{hero}

 We detected 31 HEROs down to $K=24.0$ in our observed field. The left and middle panels of Fig.\ref{fig_skyheros} show the sky distribution of the HEROs. The figure also shows the 2 $\sigma$ density contour of LAEs. It is found that the number density of HEROs in SSA22-M6 is the highest (12 HEROs) among the six fields in SSA22. In the right panel of Fig.\ref{fig_skyheros}, the surface density of HEROs is shown as a function of a radial distance from LABs, LAEs, and AzTEC sources (see the next section). The mean number density of $0.28$ arcmin$^{-2}$ in the SSA22 field is 2.3 times larger than that in GOODS-N ($0.12 \pm 0.03$ arcmin$^{-2}$) while those in SSA22-M6 is the 3.1 times larger. In addition, the marginal excesses of HEROs are found around LABs/LAEs, while there are no HEROs around LAB1 and LAB2.

 Out of the 23 HEROs detected in $R$-band (see section \ref{heroselection}), 11 have $z_{\rm phot}=2.6-3.6$. The result of the SED fitting for the 11 HEROs shows that the stellar mass range from $2.7 \times 10^{10}$ M$_{\odot}$ to $3.7 \times 10^{11}$ M$_{\odot}$, and the median is 1.6 $\times 10^{11}$ M$_{\odot}$. The 55 \% (6/11) of the HEROs are massive galaxies with $> 10^{11}$ M$_{\odot}$. It is likely that HEROs are among the most massive populations in the protocluster.

\begin{figure*}[htbp]
\includegraphics[width=6cm]{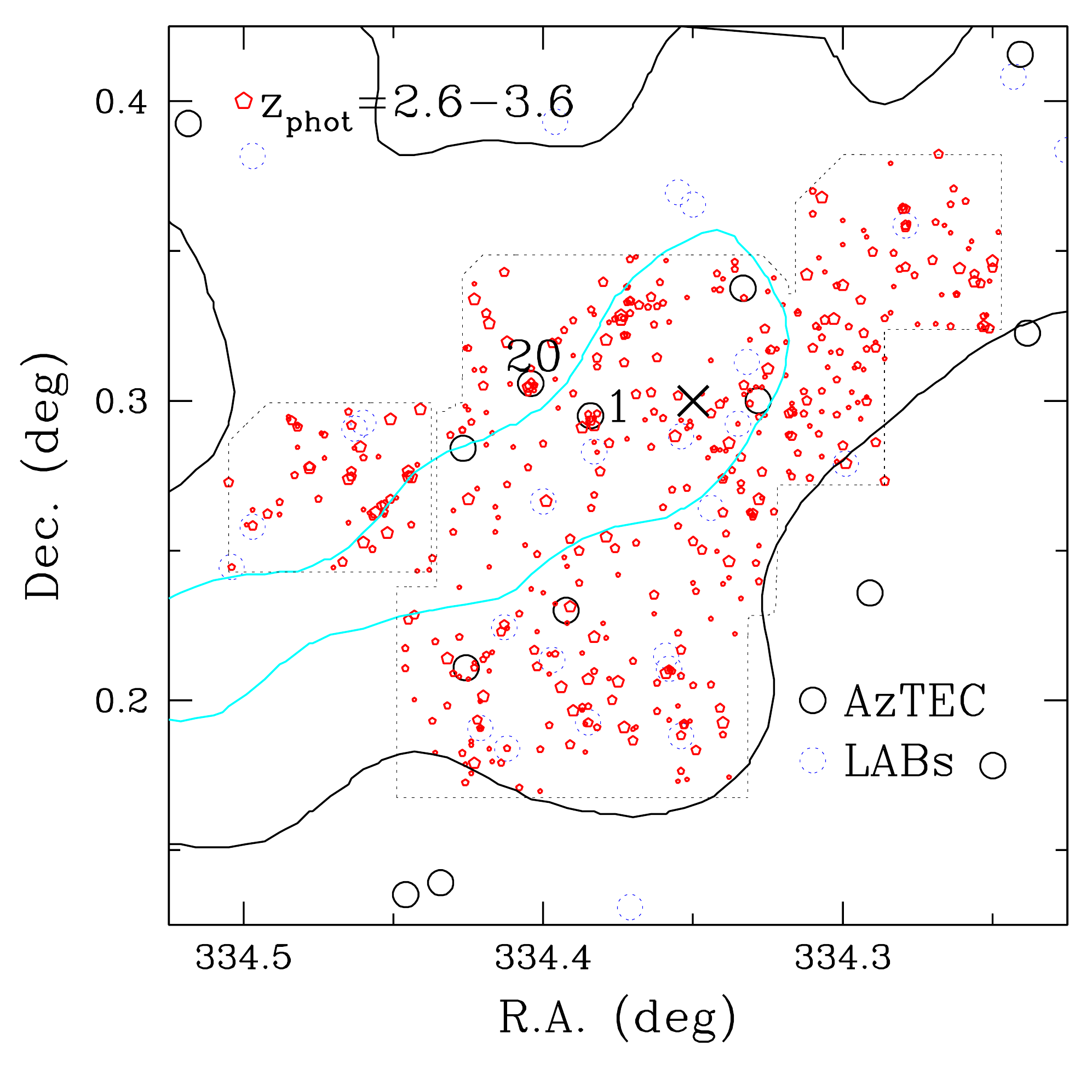}
\includegraphics[width=6cm]{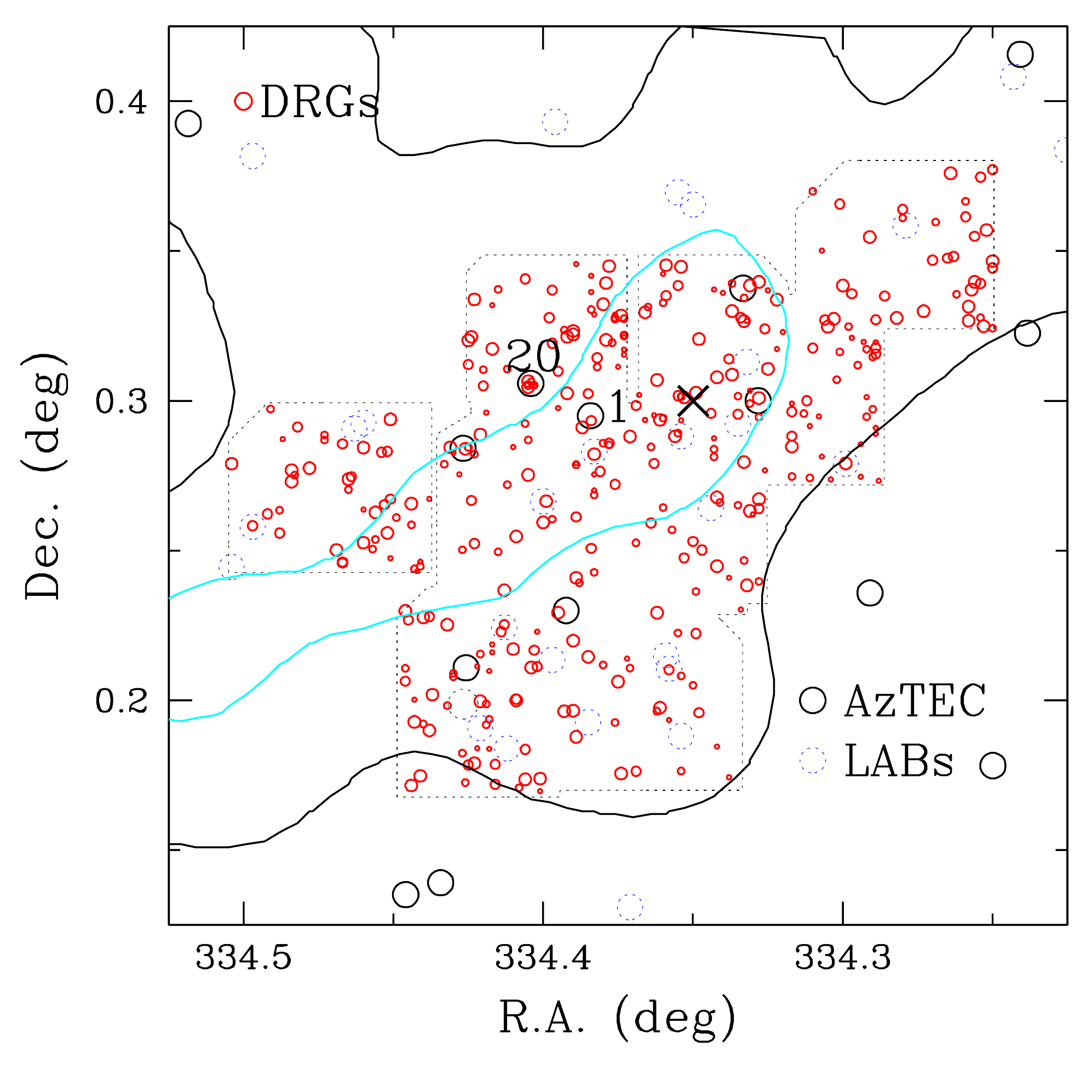}
\includegraphics[width=6cm]{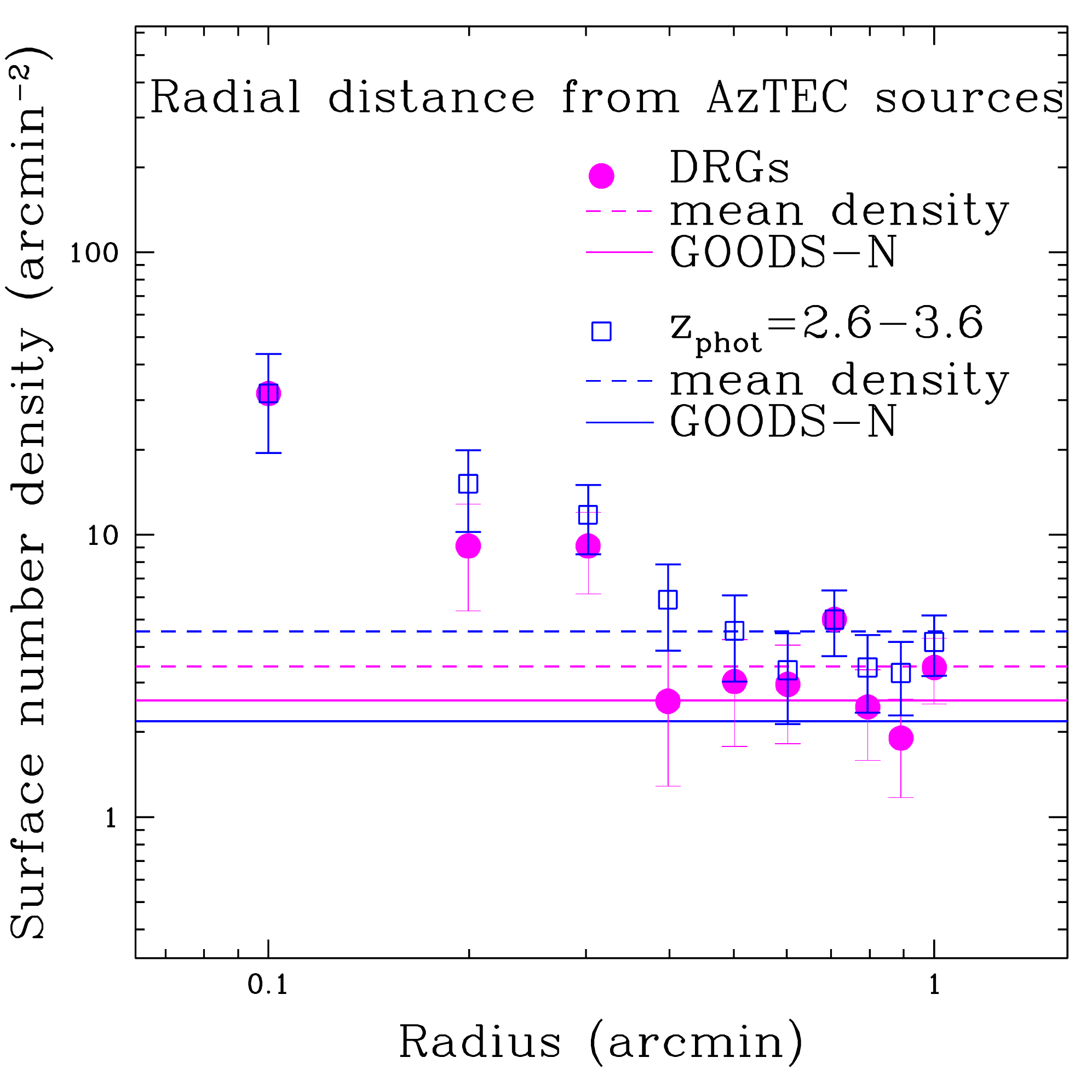}
\caption{(left)  Sky distribution of photo-z selected objects and AzTEC sources. (middle) Sky distribution of DRGs and AzTEC sources. Photo-z selected objects and DRGs are indicated with open circles (red) and filled circles (red), respectively. AzTEC sources and LABs are indicated with large open circles (black) and dotted, large open circles (blue), respectively. AzTEC1 and AzTEC20 (numbered in Tamura et al. 2009) are shown with ``1'' and ``20'', respectively. 
(right) The surface density of DRGs  (filled circles) and objects with $z_{\rm phot}=2.6-3.6$ (open squares) in a circular ring of 0.1 arcmin width as a function of radial distance from the AzTEC sources. The mean densities in GOODS-N are also indicated by the solid lines, and those in SSA22 by the dashed lines.
}
\label{fig_skyaztec}
\end{figure*}

 Of all the 31 HEROs, two objects are associated with Ly$\alpha$ emission, which are expected to be indeed located at $z=3.1$. They are found in LAB14 and the extended LAE in the vicinity of LAB35 (The object A; see section \ref{LABnoK}). Both of them are located in the 2$\sigma$ high density contour of LAEs. The object A in vicinity of LAB35 has $J-K=2.45$ and $z_{\rm phot}=3.06$. The stellar mass of this HERO is $2.65 \times 10^{11}$ M$_{\odot}$, which is one of the most massive galaxies in this region.

\subsubsection{Correlation between $K$-selected objects and AzTEC sources }
\label{aztec}

 The SSA22 protocluster is observed at 1100 $\mu$m using AzTEC camera mounted on ASTE \citep{key-tamura09}. They reported the concentration of submillimeter-bright galaxies (SMGs) in this region. Fig.\ref{fig_skyaztec} shows the sky distribution of the $K$-selected objects as well as the AzTEC sources in Tamura et al. (2009). The left panel indicates the distribution of the photo-z selected objects, and the middle one indicates that of DRGs. As shown in the right panel of Fig.\ref{fig_skyaztec}, the association between $K$-selected objects and the AzTEC sources is notable.

\begin{figure}[tbp]
\includegraphics[width=8cm]{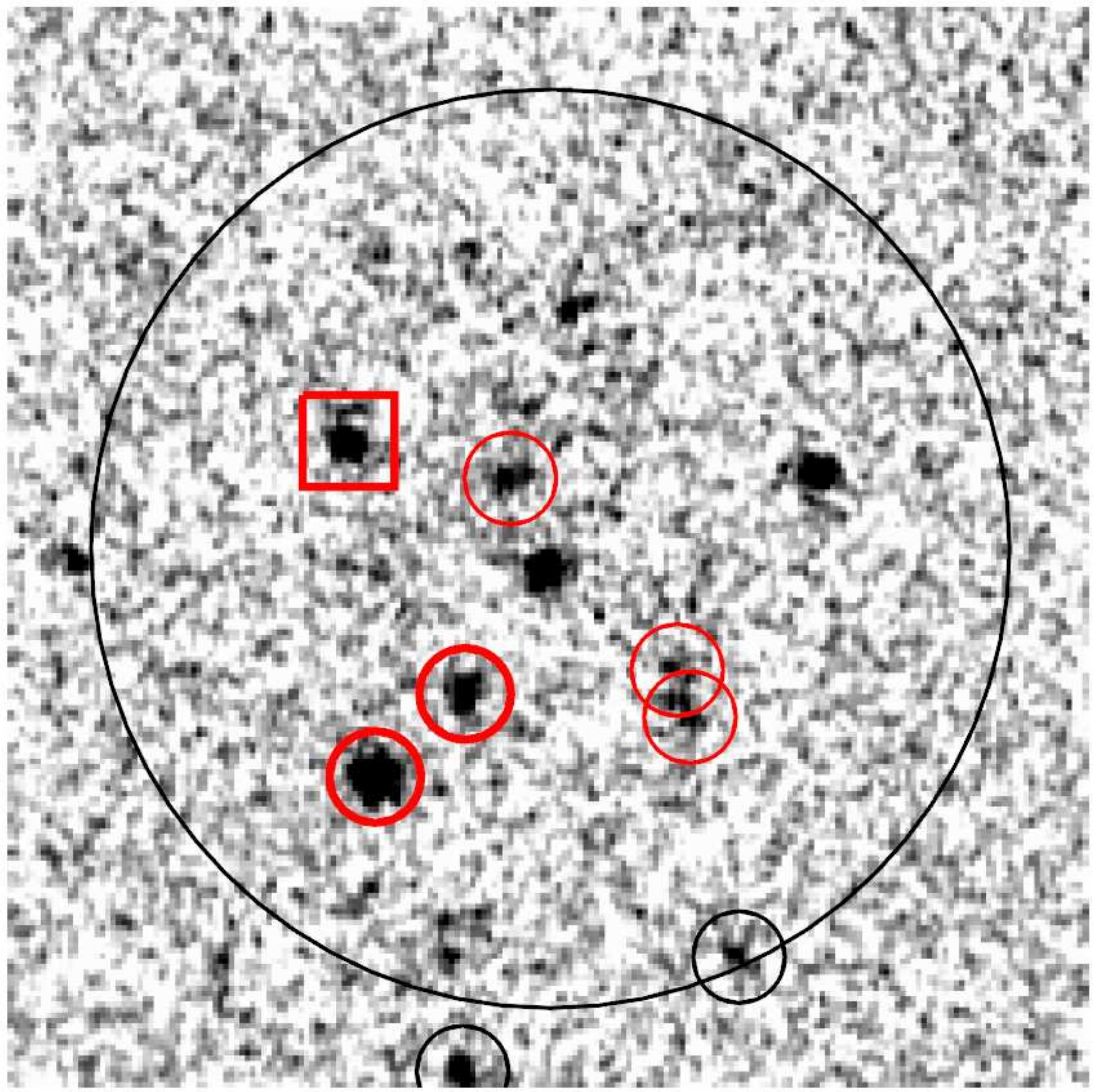}
\caption{
The $K$-selected objects around AzTEC20. The circles (black or red) indicate the photo-z selected objects with $z_{\rm phot}=2.6-3.6$. DRGs are colored red. The two thick red circles are HEROs with $z_{\rm phot}=2.6-3.6$. The red square is a HERO with $z_{\rm phot}$ slightly lower than 2.6. The large circle shows the 2$\sigma$ error range of the position of AzTEC20. The size of the image is $23''.4$ on a side.
}
\label{fig_aztec20}
\end{figure}

 Indeed, \citet{key-tamura10} reported the multi-wavelength identification of the brightest source, SSA22-AzTEC1. The object is identified with the VLA radio source, whose counterpart is detected in $K$- and $H$-band images. On the other hand, no counterpart is detected in $J$-band and optical bands. The object is classified as a HERO. \citet{key-tamura10} estimated the redshift as $z_{\rm phot}=3.19$ by their radio-submm photo-z.

 In the right panel of Fig.\ref{fig_skyheros}, we show the correlation between HEROs and the AzTEC sources. It is, however, found to be dominated by the contribution of $K$-selected objects around SSA22-AzTEC20 as shown in Fig.\ref{fig_aztec20}. They are supposed to be the counterparts of AzTEC20 and also in the ``hierarchical multiple merging'' phase. It is interesting to note that significant Ly$\alpha$ emission is not detected around the objects.

 Thus, the statistical evidence (clustering) as well as a few notable individual examples implies that at least a fraction of the AzTEC sources are indeed associated with the SSA22 protocluster.

\section{CONCLUSION}

 We present the results of wide-field deep NIR imaging in the $z=3.1$ protocluster region. The observed field is 112 arcsec$^2$ in the SSA22-Sb1 field. We extracted the candidates of the protocluster members from the $K$-selected sample by using the multi-band photo-z selection as well as the color cut for DRGs, and examined their sky distributions.

 The surface number density of DRGs in our observed field shows an excess compared with those in the blank fields, which  implies that progenitors of massive galaxies are growing in the protocluster at $z=3.1$. Interestingly, the densest area corresponds to the density peak of LAEs, and the density is 2 times higher than that in the GOODS-North field, while most of the DRGs are not directly associated with the LAEs. On the other hand, the overdensity of DRGs in the SSA22-M1 field, where is characterized by the density excess of LABs, is not significant.

 We also investigated the near-infrared properties of LABs. $K$-band counterparts with $z_{\rm phot} \simeq 3.1$ are detected for 75 \% (15/20) of the LABs within their Ly$\alpha$ halo, and the 40 \% (8/20) of LABs have multiple components. The stellar mass of 15 LABs range from $4 \times 10^{9}$ M$_{\odot}$ to $2 \times 10^{11}$ M$_{\odot}$. We also found that the stellar mass of LABs correlates with their luminosity, isophotal area, and the Ly$\alpha$ velocity widths, implying that the physical scale and the dynamical motion of Ly$\alpha$ emission are closely related to their previous star-formation activities. The results suggest that the formation of massive galaxies is in progress in a large fraction of the LABs in the field.

 Overdensity of dusty starburst galaxies such as hyper extremely red objects (HEROs; $J-K_{AB}>2.1$) and plausible $K$-band counterparts of submillimeter sources are also found in this region. Highly dust-obscured galaxies which are expected to be growing up to massive galaxies may also be populated in such high density region of Ly$\alpha$ galaxies.

 These results suggest that we are witnessing a protocluster in which the progenitors of massive galaxies are formed in the high-density region of star-forming galaxies.

\vspace*{5mm}

 We thank Dr. Yoichi Tamura for providing the updated coordinates of the AzTEC sources. We thank the staff of Subaru Telescope for their assistance with the development and the observation of MOIRCS. This research is supported in part by the Grant-in-Aid 20450224 for Scientific Research of the Ministry of Education, Science, Culture, and Sports in Japan. The Image Reduction and Analysis Facility (IRAF) used in this paper are distributed by National Optical Astronomy Observatories, U.S.A., operated by the Association of Universities for Research in Astronomy, Inc., under contact to the U.S.A. National Science Foundation.




\clearpage

\begin{deluxetable}{ccccccc}
\tablewidth{0pt}
\tabletypesize{\footnotesize}
\tablecaption{Summary of the observations \label{tab_obs}} 
\tablehead{
\colhead{Field} & 
\multicolumn{2}{c}{Center position} & 
\colhead{Filter} & 
\colhead{Date} & 
\colhead{Seeing} & 
\colhead{Exposure}\\
\colhead{ (size)} & 
\colhead{R.A.} & 
\colhead{Dec.} & 
\colhead{} & 
\colhead{(HST)} & 
\colhead{(arcsec)} & 
\colhead{(hours)} 
}
\startdata
SSA22-M1 & 22:17:33.9 & 00:12:14.0 & $J$  & Jun. 15, 2005    & 0.53 & 0.9 \\
($7' \times 4'$) &   &  & $H$  & Aug. 14, 2005, Jul. 22, 2006    & 0.48 & 1.0 \\
                    &  & & $K_{\rm s}$  & Jun. 14, 15, Aug. 14, 2005  & 0.43 & 1.5 \\
    &  & & & Jul. 22, 2006  &  &  \\
\\
SSA22-M2 & 22:17:31.9 & 00:15:44.0 & $J$  & Jul. 22, 2006    & 0.51 & 0.9 \\
($7' \times 4'$) &              &  & $H$  & Jul. 22, 2006    & 0.47 & 0.6 \\
    &  & & $K_{\rm s}$                   & Jul. 22, 2006, Sep. 24, 2007 & 0.47 & 1.1 \\
\\
SSA22-M3 & 22:17:16.6 & +00:17:58.7  & $J$  & Sep. 22, 2007    & 0.48 & 2.0 \\
($4'  \times 3'.5  $) &  &  & $H$       & Sep. 22, 23, 2007    & 0.59 &  1.7 \\
    &  & & $K_{\rm s}$  & Sep. 22, 2007  & 0.51                               &  1.2 \\
\\
SSA22-M4 & 22:17:52.8 & +00:16:15.4 & $J$  & Sep. 22, 23, 2007    & 0.60 & 2.0 \\
 ($4'  \times 3'.5  $)&  &  & $H$  &  Sep. 23, 2007   & 0.50             & 1.6 \\
    &  & & $K_{\rm s}$  & Sep. 22, 23, 2007  & 0.53                              & 1.5 \\
\\
SSA22-M5 & 22:17:07.4 & +00:21:07.5 & $J$  & Sep. 24, 2007    & 0.67 & 2.1 \\
($4'  \times 3'.5  $) &  &  & $H$  & Sep. 24, 2007    & 0.46         & 1.1 \\
    &  & & $K_{\rm s}$  & Sep. 23, 24, 2007  & 0.47                          & 1.3 \\
\\
SSA22-M6 & 22:17:28.7 & +00:19:07.5 & $J$  & May 16, 17, 2008    & 0.47 & 1.2 \\
($7' \times 4'$) &  &  & $H$  &  May 17, 2008    & 0.54                 & 0.6 \\
    &  & & $K_{\rm s}$  &  May 16, 2008  & 0.41                                 & 1.0 \\
\enddata
\end{deluxetable}       

 
\clearpage

\begin{deluxetable}{lccccccccc} 
\tablewidth{0pt}
\tabletypesize{\footnotesize}
\tablecaption{The limiting magnitude in each field \label{tab_limitmag}} 
\tablehead{
\colhead{Observed Area} & 
\colhead{$K^a$} &  
\colhead{$J^b$} & 
\colhead{$K^b$} &  
\colhead{$J^c$} & 
\colhead{$H^c$} & 
\colhead{$K^c$} &  
\colhead{$J^d$} & 
\colhead{$H^d$} & 
\colhead{$K^d$}  \\
\colhead{} & 
\colhead{($1'' \phi$, 5$\sigma$)} & 
\multicolumn{2}{c}{($1.1'' \phi$, 5$\sigma$) } & 
\multicolumn{3}{c}{($1.4'' \phi$, 5$\sigma$) } & 
\multicolumn{3}{c}{($2.0'' \phi$, 5$\sigma$) } 
}

\startdata
SSA22-M1 & 24.4  & 24.6 & 24.5 &  24.4  & 24.0 & 24.1 & 23.9 & 23.5 & 23.6 \\ 
SSA22-M2 & 24.3  & 24.6 & 24.2 &  24.2  & 24.0 & 23.9 & 23.6 & 23.5 & 23.4 \\ 
SSA22-M3 & 24.3  & 25.0 & 24.3 &  24.7  & 24.0 & 24.1 & 23.2 & 23.6 & 23.6 \\ 
SSA22-M4 & 24.3  & - & - &  24.3  & 24.2 & 23.8 & 23.6 & 23.5 & 23.3 \\
SSA22-M5 & 24.2  & - & - &  24.2  & 23.8 & 23.9 & 23.9 & 23.3 & 23.5 \\ 
SSA22-M6 & 24.3  & 24.7 & 24.3 &  24.3  & 23.8 & 24.0 & 23.7 & 23.4 & 23.5 \\ 
\enddata

\tablenotetext{a}{Measured in original images with seeing FWHM in Table \ref{tab_obs}.}
\tablenotetext{b}{Measured in smoothed images with $0''.5$ FWHM.}
\tablenotetext{c}{Measured in smoothed images with $0''.7$ FWHM.}
\tablenotetext{d}{Measured in smoothed images with $1''.0$ FWHM.}

\end{deluxetable}       

\clearpage

\begin{deluxetable}{rcccclll} 
\tablecolumns{8}
\tablewidth{356pt}
\tabletypesize{\footnotesize}
\tablecaption{Near-infrared properties of LABs \label{tab_labNIR}} 
\tablehead{
\colhead{LAB} &  
\colhead{ $K$ } & 
\colhead{ err } & 
\colhead{ $J-K$ } & 
\colhead{ err } & 
\colhead{ $z_{\rm phot}$ } &
\colhead{ $z_{\rm spec}$ } & 
\colhead{ Comment }
}
\startdata

LAB1 & & & & & & & \\
1 & 22.02 & 0.10 & 1.24 & 0.23 & 3.38  & &\\ 
2 & 22.97 & 0.11 & 1.47 & 0.34 & 3.08  & 3.097 & DRG\\ 
3 & 23.34 & 0.11 & 1.15 & 0.31 & 3.18  & 3.109 & LBG C11$^a$\\ 
4 & 23.69 & 0.19 & $>$1.08 & - & 3.35  & &\\ 
5 & 23.97 & 0.22 & 0.76 & 0.51 & 0.05  & &\\ 
6 & 24.57 & 0.23 & $>$0.78 & -& 2.96  & &\\ 
LAB2 & & & & & & & \\
1 & 22.56 & 0.10 & 2.07 & 0.38 & 3.04  & & DRG\\ 
2 & 22.91 & 0.11 & 1.68 & 0.42 & 3.00  & (3.091) & DRG, vicinity of LBG M14$^b$ \\ 
3 & 23.68 & 0.20 & $>$1.03 & - & 3.25  & &\\ 
4 & 24.30 & 0.14 & 0.41 & 0.52 & 2.98  & &\\ 
5 & 24.51 & 0.26 & 0.75 & 0.66 & 3.65  & &\\ 
LAB3 & & & & & & & \\
1 & 22.54 & 0.12 & 1.47 & 0.29 & 3.10  & & DRG\\ 
LAB5 & & & & & & & \\
1 & 22.33 & 0.10 & 1.61 & 0.19 & 2.65  & & DRG\\ 
LAB7 & & & & & & & \\
1 & 23.49 & 0.12 & $>$1.26 & - & 3.19  & 3.098 & DRG, LBG C6 \\ 
2 & 23.65 & 0.13 & 0.19 & 0.26 & 2.65  & 3.093 & LBG M4 \\ 
3 & 23.97 & 0.18 & 0.71 & 0.61 & 3.35  & (3.098) & (LBG C6), vicinity of 1	\\ 
LAB8 & & & & & & & \\
3 & 24.35 & 0.28 & 0.66 & 0.47 & 3.35  & 3.094 & LBG C15	\\ 
LAB11 & & & & & & & \\
1 & 23.79 & 0.17 & 0.65 & 0.33 & 2.79  & 3.0748 & LBG C47	\\ 
2 & 24.49 & 0.23 & 1.07 & 0.47 & 2.71  & (3.0748) & LBG C47,	vicinity of 1	\\ 
3 & 24.94 & 0.24 & 0.59 & 0.70 & 3.35  & (3.0748) & LBG C47,	vicinity of 1	\\ 
LAB12 & & & & & & & \\
1 & 22.11 & 0.10 & 2.14 & 0.45 & 2.12  & & DRG \\ 
2 & 23.67 & 0.21 & 1.16 & 0.35 & 3.40  & 3.094 & LBG M28\\ 
3 & 24.33 & 0.18 $>$& 1.29 & - & 3.36  & (3.094) & (LBG M28), vicinity of 2\\ 
LAB14 & & & & & & & \\
1 & 22.45 & 0.11 & 1.81 & 0.34 & 3.10  & 3.089 & DRG \\ 
LAB16 & & & & & & & \\
1 & 22.99 & 0.11 & 0.92 & 0.24 & 2.71  & &\\ 
2 & 24.22 & 0.23 & $>$1.11 & - & 3.35  & &\\ 
LAB20 & & & & & & & \\
1 & 23.30 & 0.11 & 0.93 & 0.36 & 2.58  & 3.118 & LBG C12	 \\ 
LAB24 & & & & & & & \\
1	  & 23.36 & 0.13 & - & - & 	3.08 &  		 & 	DRG	\\
LAB27 & & & & & & & \\
1 & 22.83 & 0.12 & 1.28 & 0.26 & 2.83  & &\\ 
2 & 22.83 & 0.12 & 0.92 & 0.34 & 2.76  & &\\ 
3 & 23.79 & 0.13 & 0.96 & 0.40 & 2.81  & &\\ 
4 & 23.96 & 0.17 & $>$0.89 & - & 3.42  & &\\ 
LAB30 & & & & & & & \\
1 & 22.96 & 0.11 & 0.79 & 0.20 & 3.02  & 3.086 & LBG D3	\\ 
2 & 23.51 & 0.19 & 0.39 & 0.31 & 3.23  & &\\ 
LAB31 & & & & & & & \\
1 & 23.20 & 0.11 & 1.36 & 0.35 & 2.87  & 3.076 & LBG C4\\ 
LAB35 & & & & & & & \\
A & 21.73 & 0.10 & 2.45 & 0.26 & 3.06  & &	DRG\\ 
B & 23.04 & 0.11 & 1.85 & 0.54 & 0.84  & &	DRG\\

\enddata
\tablenotetext{a}{ID of the LBG in Steidel et al. (2003)}
\tablenotetext{b}{$0''.9$ apart from the rest-frame UV source}
\end{deluxetable}       

\clearpage

\begin{deluxetable}{lcccccccl} 
\tablecolumns{8}
\tablewidth{0pt}
\tabletypesize{\footnotesize}
\tablecaption{The results of the SED fitting of photo-z selected objects in LABs \label{tab_labSED}} 
\tablehead{
\colhead{LAB} &  
\colhead{ID } & 
\colhead{ $M_*$ } & 
\colhead{ $\chi ^2$ } & 
\colhead{ Mv}\\
\colhead{} &  
\colhead{ } & 
\colhead{ ($10^{10} M_{\odot}$) } &
\colhead{  } & 
\colhead{ }}

\startdata
 LAB1 
  &     1       & 14.49  $_{-7.69} ^{+4.96}$ &   0.70 & -23.01 \\
  &     2       &  9.15  $_{-5.83} ^{+5.02}$ &   0.39 & -22.52 \\
  &     3       &  0.80  $_{-0.71} ^{+2.39}$ &   0.28 & -21.95 \\
  &     4       &  2.41  $_{-2.15} ^{+3.41}$ &   0.07 & -21.67 \\
  &     5       &  0.38  $_{-0.24} ^{+5.66}$ &   0.10 & -21.68 \\
  &     6       &  1.14  $_{-1.12} ^{+1.69}$ &   0.12 & -21.40 \\
 LAB2 
  &     1       &  6.81  $_{-3.73} ^{+55.73}$ &  0.03 & -22.91 \\
  &     2       & 10.92  $_{-10.05} ^{+10.25}$ & 0.20 & -22.64 \\
  &     3       &  4.64  $_{-4.44} ^{+5.43}$ &   0.05 & -21.80 \\
  &     4       &  1.44  $_{-1.33} ^{+1.52}$ &   0.14 & -21.65 \\
  &     5       &  3.72  $_{-3.67} ^{+9.84}$ &   0.01 & -21.31 \\
 LAB3 
  &     1       &  9.26  $_{-4.28} ^{+7.30}$ &   0.18 & -22.93 \\
 LAB5 
  &     1       &  7.19  $_{-2.55} ^{+2.50}$ &   0.99 & -23.27 \\
 LAB7 
  &     1       &  0.82  $_{-0.74} ^{+2.26}$ &   0.12 & -21.72 \\
  &     2       &  0.75  $_{-0.68} ^{+2.47}$ &   0.18 & -21.73 \\
  &     3       &  0.60  $_{-0.58} ^{+1.62}$ &   0.06 & -21.31 \\
 LAB8 
  &     1       &  0.42  $_{-0.41} ^{+2.32}$ &   0.01 & -20.98 \\
 LAB11 
  &     1       &  0.42  $_{-0.39} ^{+1.35}$ &   0.05 & -21.47 \\
  &     2       &  0.57  $_{-0.55} ^{+3.29}$ &   0.07 & -21.30 \\
  &     3       &  0.58  $_{-0.52} ^{+1.65}$ &   0.08 & -21.84 \\
 LAB12 
  &     1       & 17.44  $_{-9.19} ^{+50.20}$ &  1.15 & -23.02 \\
  &     2       &  2.07  $_{-1.86} ^{+1.93}$ &   0.33 & -21.94 \\
  &     3       &  1.44  $_{-1.37} ^{+1.88}$ &   0.31 & -21.65 \\
 LAB14 
  &     1       &  16.95  $_{-9.91} ^{+5.19}$ &  0.10 & -22.95 \\
 LAB16 
  &     1       &  1.76  $_{-1.29} ^{+3.82}$ &   0.55 & -22.48 \\
  &     2       &  0.95  $_{-0.92} ^{+1.50}$ &   0.03 & -21.33 \\
 LAB20 
  &     1       &  1.56  $_{-1.29} ^{+1.39}$ &   0.17 & -21.86 \\
 LAB24 
  &     1       &  2.85  $_{-1.69} ^{+2.78}$ &   0.24 & -21.91 \\
 LAB27 
  &     1       &  1.77  $_{-1.16} ^{+1.97}$ &   0.85 & -22.56 \\
  &     2       &  3.32  $_{-2.85} ^{+2.08}$ &   0.31 & -22.20 \\
  &     3       &  0.38  $_{-0.34} ^{+1.44}$ &   0.25 & -21.56 \\
  &     4       &  1.77  $_{-1.70} ^{+3.01}$ &   0.14 & -21.63 \\
 LAB30 
  &     1       &  1.72  $_{-1.02} ^{+2.14}$ &   0.10 & -22.56 \\
  &     2       &  1.34  $_{-1.25} ^{+2.99}$ &   0.19 & -21.80 \\
 LAB31 
  &     1       &  2.56  $_{-1.59} ^{+1.97}$ &   0.24 & -22.17 \\
 (LAB35) 
  &     A       & 26.52 $_{-11.69} ^{+38.37}$ &  0.12 & -23.74 \\

\enddata
\end{deluxetable}       

\end{document}